\documentclass[sn-mathphys,Numbered]{sn-jnl}

\usepackage{graphicx}%
\usepackage{multirow}%
\usepackage{amsmath,amssymb,amsfonts}%
\usepackage{geometry}
\usepackage{amsthm}%
\usepackage{mathrsfs}%
\usepackage[title]{appendix}%
\usepackage{xcolor}%
\usepackage{textcomp}%
\usepackage{manyfoot}%
\usepackage{booktabs}%
\usepackage{algorithm}%
\usepackage{algorithmicx}%
\usepackage{algpseudocode}%
\usepackage{listings}%
\usepackage{geometry}%
\usepackage{ulem}
\usepackage{color}
\usepackage{cite}

\theoremstyle{thmstyleone}%
\theoremstyle{thmstyletwo}%

\theoremstyle{thmstylethree}%

\begin{document}

\title[Article Title]{Optical Spiking Neurons Enable High-Speed and Energy-Efficient Optical Neural Networks}

\author[1]{\fnm{Bo} \sur{Xu}}\email{bxu721@connect.hkust-gz.edu.cn}
\equalcont{These authors contributed equally to this work.}

\author[1]{\fnm{Zefeng} \sur{Huang}}\email{24040193r@connect.polyu.hk}
\equalcont{These authors contributed equally to this work.}

\author[1]{\fnm{Yuetong} \sur{Fang}}
\author[1]{\fnm{Xin}\sur{Wang}}
\author[1]{\fnm{Bojun}\sur{Cheng}}
\author*[2]{\fnm{Shaoliang}\sur{Yu}} \email{yusl@zhejianglab.org}
\author*[3]{\fnm{Zhongrui}\sur{Wang}}\email{wangzr@sustech.edu.cn}
\author*[1]{\fnm{Renjing} \sur{Xu}}\email{renjingxu@hkust-gz.edu.cn}

\affil[1]{\orgdiv{Thrust of Microelectronics of Function Hub}, \orgname{The Hong Kong University of Science and Technology (Guangzhou)}, \orgaddress{\city{Nansha}, \postcode{511400}, \state{Guangzhou}, \country{China}}}

\affil[2]{\orgdiv{Zhejiang Laboratory}, \orgaddress{\city{Hangzhou}, \postcode{311100}, \country{China}}}

\affil[3]{\orgdiv{School of Microelectronics}, \orgname{Southern University of Science and Technology}, \orgaddress{\city{Shenzhen}, \country{China}}}

\abstract{
Optical neural networks (ONNs) perform extensive computations using photons instead of electrons, resulting in passively energy-efficient and low-latency computing. Among various ONNs, the diffractive optical neural networks (DONNs) particularly excel in energy efficiency, bandwidth, and parallelism, therefore attract considerable attention. However, their performance is limited by the inherent constraints of traditional frame-based sensors, which process and produce dense and redundant information at low operating frequency. Inspired by the spiking neurons in human neural system, which utilize a thresholding mechanism to transmit information sparsely and efficiently, we propose integrating a threshold-locking method into neuromorphic vision sensors to generate sparse and binary information, achieving microsecond-level accurate perception similar to human spiking neurons. By introducing novel Binary Dual Adaptive Training (BAT) and Optically Parallel Mixture of Experts (OPMoE) inference methods, the high-speed, spike-based diffractive optical neural network (S\textsuperscript{2}NN) demonstrates an ultra-fast operating speed of 3649 FPS, which is 30 fold faster than that of reported DONNs, delivering a remarkable computational speed of 417.96 TOPS and a system energy efficiency of 12.6 TOPS/W. Our work demonstrates the potential of incorporating neuromorphic architecture to facilitate optical neural network applications in real-world scenarios for both low-level and high-level machine vision tasks.

}

\maketitle

\section{Main}\label{sec1}

Optical computing has garnered significant research interest for decades, particularly in light of the exponential growth in computational and latency demands of modern deep neural networks (DNNs), which have outpaced Moore's law improvements in microprocessor performance~\citep{bib1,bib3,bib30,bib31,bib48,bib50,bib64}. Photons, unlike electrons, essentially propagate without generating heat and suffering from signal degradation due to inductive and capacitive effects~\citep{bib2,bib51,bib52}. Endowed with photonics inherent high-speed, high-energy efficiency and densely parallel characteristics, optical neural networks (ONNs) has been heralded as a prospective solution to alleviate the limitations of electronic computing~\citep{bib6,bib7,bib32,bib33,bib34,bib60,bib61,bib63}. 

DONNs have garnered considerable attention due to their low control complexity and high scalability among optical computing approaches~\citep{bib8,bib9,bib10,bib35,bib36,bib37,bib59}. For large integrated photonics architectures based on the MZI mesh or MRR array, the cumulative impact of systematic errors, induced by factors such as random phase noise~\citep{bib54}, thermal crosstalk~\citep{bib54}, thermal fluctuations and fabrication imperfections~\citep{bib53}, becomes increasingly problematic and could significantly compromise the scalability and application of integrated photonics~\citep{bib11,bib12,bib13,bib14,bib38}. In contrast, free-space diffractive optical systems, which integrate digital micromirrors devices (DMD), spatial light modulators (SLM), and frame-based sensors, not only provide greater flexibility and additional spatial degrees of freedom that potentially mitigate these limitations~\citep{bib2}, but also offer greater bandwidth and parallelism due to their seamlessly scalable architecture. In this line of research, DONNs have been constructed for visual processing and all-optical computational tasks such as depth detection~\citep{bib15}, image classification \citep{bib4}, image sensing~\citep{bib62}, saliency detection~\citep{bib15}, and human action recognition~\citep{bib16}. Although DONNs exploit the merit of low-loss, parallel, ultrafast, and clock-free light propagation, the dense and float information produced by the receiver (frame-based sensors) pose severe limitations to inference speed~\citep{bib16}. In addition, frame-based senors have a high energy consumption, high information redundancy, and apply only in limited ranges and environments~\citep{bib17}. The resulting interframe processing interval for frame-based DONN architecture is normally larger than 30 ms, which is insufficient for advanced machine learning tasks~\citep{bib18}. 

Compared to traditional frame-based sensors, the human visual nervous system operates on a fundamentally different principle by encoding visual information through sparse spikes and decoding it in the brain, thus efficiently acquiring information~\citep{bib24,bib25,bib26}. Building on this biological paradigm, we constructed optical spiking neurons (OSNs) based on a threshold-locking method that leverages neuromorphic vision sensors to achieve humanoid-like perceptual capabilities. Neuromorphic vision sensors generate data asynchronously only when capturing changes in pixel-level illumination intensity, rather than producing stable frames at fixed time intervals. This unique principle provides advantages such as low motion blur, high dynamic range, spatio-temporal sparsity, and microsecond-level resolution, all while using less bandwidth and power resources~\citep{bib19,bib39,bib40,bib49}. Despite these benefits, previous research has often treated neuromorphic vision sensors merely as initial inputs in a processing pipeline or as supplements to traditional frame-based sensors, overlooking their distinctive neuromorphic properties~\citep{bib19,bib20,bib21,bib22,bib23}, which are crucial for streamlining large systems with bulky components and reducing overall energy consumption. However, in order to fully leverage the advantages of neuromorphic vision sensors,
the first issue that needs to be addressed is the temporal correlation in their outputs. 
Due to the inherent limitations of neuromorphic vision sensors, it is inevitable that previous outputs will influence current ones. Thus, it will lead to information crosstalk during high-speed operations, despite their speed and energy efficiency. This issue also poses a challenge for neuromorphic vision sensors to function directly as neurons in DONNs systems (Supplementary Note 1).

To address this issue, in this study, we introduced a method of optical threshold-locking to perform a hard reset after each sensing process in the neuromorphic vision sensor, resolving the problem of information crosstalk. Based on this method, we designed and implemented truly functional OSNs and fully explore their intrinsic brain-inspired characteristics. By redesigning the diffractive processing unit (DPU) to integrate with OSNs, we developed a high-speed spike-based diffractive neural network (S\textsuperscript{2}NN) for advanced visual machine learning tasks. Specifically: 
(i) Optical spiking neurons in the sDPU emulate biological neuronal activities with firing and resetting mechanisms, using a threshold-locking method to prevent crosstalk and ensure binary and sparse pulse outputs during high-speed information transmission, thereby enabling spike-based diffractive process unit (sDPU) based optoelectronic systems to transmit vast amounts of refined information within microseconds and significantly increase the system's inference speed.
(ii) We established a comprehensive architecture that supports both training and inference phases, significantly enhancing the parametric optimization of the physical system and achieving more than a 30-fold increase in inference speed compared to previous systems, reaching up to 3649 FPS. (iii) To accommodate video inputs, we also developed the temporal-stacked S\textsuperscript{2}NN (S\textsuperscript{3}NN) and single-shot S\textsuperscript{2}NN (S\textsuperscript{4}NN) models to meet the demands of advanced applications while maintaining high-speed dynamics features. The S\textsuperscript{3}NN model reduces redundant information at the input end, whereas the S\textsuperscript{4}NN model processes outputs using Optically Parallel Mixture of Experts (OPMoE) principles to achieve higher inference speeds. Our OSN-based system demonstrated superior performance on the Modified National Institute of Standards and Technology (MNIST) dataset~\citep{bib27}, human action recognition datasets (KTH and Weizmann)~\citep{bib28, bib29}. Compared to existing diffractive photonic computing methods~\citep{bib2,bib15}, our OSN-based system improves efficiency and reduces inference time by more than an order of magnitude. Different networks have been validated to achieve advanced machine learning tasks within microsecond time frames. The OSN architecture enhances high-performance photonic neural networks and enables real-time analysis of dynamic visual scenes beyond microsecond time scales.

\section{Optical spiking neuron}\label{sec2}

The principle and optoelectronic implementation of the proposed sDPU are illustrated in Fig.~\ref{fig1}a, with forward model details in Supplementary Note 2. Compared to traditional DPU, the neuromorphic vision sensor acts as OSNs, generating binary and sparse output by applying a complex activation function to the calculated optical field that naturally occurs during photoelectric conversion.

In our proposed architecture, a high-speed programmable optoelectronic system performs data routing, replication, weighting, summation, and activation with a fast frame time, on the order of microseconds, as shown in Fig.~\ref{fig1}a, Fig.~\ref{fig1}b and Supplementary Fig. S1. The DMD encodes the quantized input into a complex-valued optical field. Different input nodes are physically connected to individual output neurons through light diffractive connections, where a reconfigurable weighting element, an SLM, controls the strength of the connections determined by the diffractive modulation of the wavefront. The weighted results are then transmitted to the neuromorphic vision sensor. In traditional setups, the generation of float and dense data by frame-based sensors leads to significant redundancy, severely impeding system performance. Thus, we propose using OSN to increase information sparsity, which coordinates high accuracy with microsecond frame speeds.

To elucidate the OSN, Fig.~\ref{fig1}c-d shows the pipeline involving a whole spiking process of a single pixel in OSN. The threshold for an OSN is set using a threshold light that could perform a hard reset. This threshold is determined through neural network training and is finely tuned to achieve results comparable to those in simulations (see Supplementary Fig. S2). When the intensity of incident light surpasses this predetermined threshold, the neuron is triggered to fire. If the intensity of incident light falls below the threshold, the neuron remains inactive. Subsequent to its activation, the neuron undergoes a hard reset induced by the threshold light again, resetting its threshold for the next incident input. Since only positive spikes will be generated, we could simply achieve binary and sparse output, which significantly reduces data redundancy and enhances both computing speed and system energy efficiency compared to cutting-edge GPU processors~\citep{bib41,bib42}.

With this setup, our OSN behavior extends from the neuromorphic vision sensor, allowing our OSN to more closely resemble the spiking neuron in the human brain. 
The original response of a neuromorphic vision sensor can be described through the response of one pixel. Each pixel dynamically stores a reference level which is set by the last timestamp when detecting intensity variations (as low as 25\% with the model we use; see Supplementary Note 1) in a logarithmic manner. Numerous spikes are returned, each providing the x and y pixel coordinates, the polarity (positive or negative), and the timestamp of the trigger. In our case, we employed a threshold light to establish a fixed threshold, which aids in mitigating the excessive generation of spike bursts, thereby yielding a binary output from the noisy spike buffer. In this proof-of-concept study, the system is configured with a base frequency gradually increased to about 2000 Hz and a average frame interval of 500 $\mu s$. The pulse repetition rate is variable, ranging from 250 Hz to about 2000 Hz, with the DMD outputting pulses of variable width corresponding to these rates. Following the fire-hard reset process, our OSN can stably and repeatedly extract binarized, sparse features, which significantly reduces information volume and energy consumption (Supplementary Note 3), thus laying the foundation for enhanced overall operational speed.  

In Fig.~\ref{fig1}e-f, we set out to elucidate the characteristics of a single OSN, focusing on its response to various incident light intensities and reset thresholds. In the experiment, the threshold optical power when the LED is on is systematically varied between 0.2 $\mu$W and 20.8 $\mu$W, corresponding to different levels from 0 to 20. Simultaneously, the laser power is adjusted from 0.33 mW to 3.3 mW and normalized to a range of 0 to 1. We first assessed the response of the OSN across four threshold levels, and our results indicated a ReLU-like response pattern~\citep{bib44}, which also assists in minimizing the inherent uncertainty of the spiking behavior~\citep{bib43}. Furthermore, we characterized how different operating frequencies and laser intensities influence the threshold necessary for reliable neuron activation, as displayed in Fig.~\ref{fig1}f. As expected, as the frequency increases, the required threshold decreases, and vice versa for the laser intensity.

With OSN, excessive information can be passively filtered and maintained at a sparse level, enabling the processing of high-speed dynamics without the need for additional CPU or GPU processing, which would otherwise hinder high bandwidth and inference speed brought by light.

\begin{figure}[H]%
\centering
\includegraphics[width=1.0\textwidth]{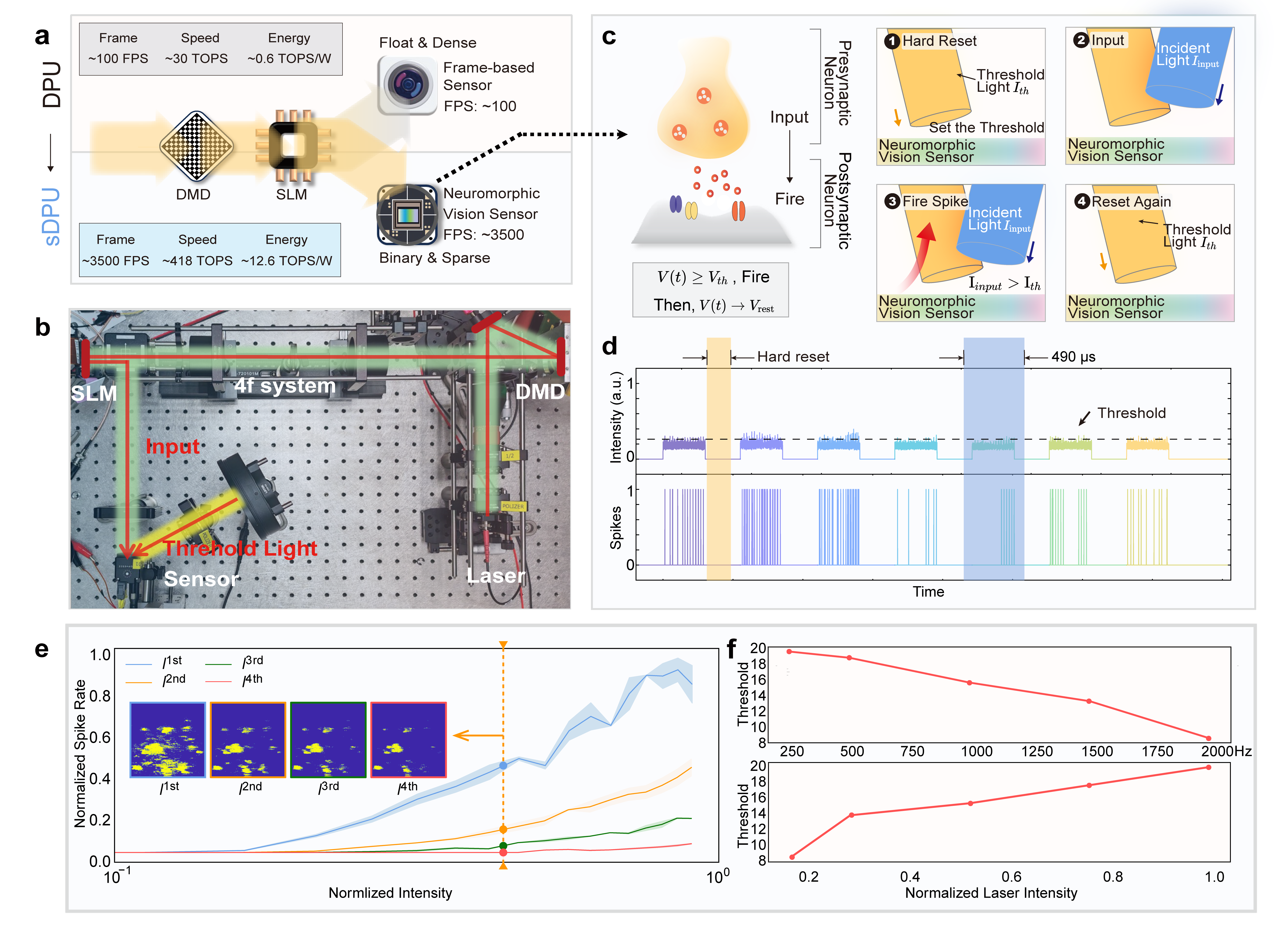}
\caption{\textbf{High-speed and energy-efficient OSN}
\textbf{a,} System composition and performance comparison of DPU and sDPU. By replacing the detector with a neuromorphic vision sensor and introducing the threshold light, the overall operating speed of sDPU is increased by approximately 30-fold compared to DPU. 
\textbf{b,} Actual optical path diagram. The actual optical path includes a laser, DMD, SLM, neuromorphic vision sensor, an external light source providing threshold light, and the corresponding optical components (detailed in Methods and Supplementary Fig. S1). 
\textbf{c,} Inspiration and implementation process of OSNs in sDPU system. The inspiration for OSNs comes from biological neurons, where input excitation greater than the resting potential leads to the release of information from the pre-synaptic to the post-synaptic end. In OSNs, the coherent light field acts as the excitation, and the threshold light serves as the resting potential. When the input meets the condition, it triggers the generation of spikes. 
\textbf{d,} Spikes output of OSNs under different inputs. The spikes output of OSNs under different inputs corresponds to the working principle in Figure c. Only light that meets certain conditions will trigger the OSN to generate spikes. 
\textbf{e,} Response characteristics of OSNs. The response characteristics of OSNs: \textit{I}\textsuperscript{1st}, \textit{I}\textsuperscript{2nd}, \textit{I}\textsuperscript{3rd}, \textit{I}\textsuperscript{4th} represent the relationship between spikes rate and input light intensity at four different threshold light intensities. With the same input light intensity, a higher threshold light intensity results in a lower spikes rate.  Inset: Different feature maps under varying intensities of threshold light. An optimal threshold can filter out most of the noise while retaining the main features of the image. 
\textbf{f,} Upper: Required threshold light intensity at different operating frequencies. At higher operating frequencies, the light intensity per unit time decreases, so the threshold must be lowered to obtain the desired output. Down: Adjusting threshold light intensity for desired output at 2000 Hz. When the input light is too strong, the threshold light intensity needs to be increased to achieve the desired output at a fixed operating frequency of 2000 Hz. 
}\label{fig1}
\end{figure}

\section{Results}\label{sec3}
\subsection{S\textsuperscript{2}NN}\label{subsec2}

\begin{figure}[H]%
\centering
\includegraphics[width=1.0\textwidth]{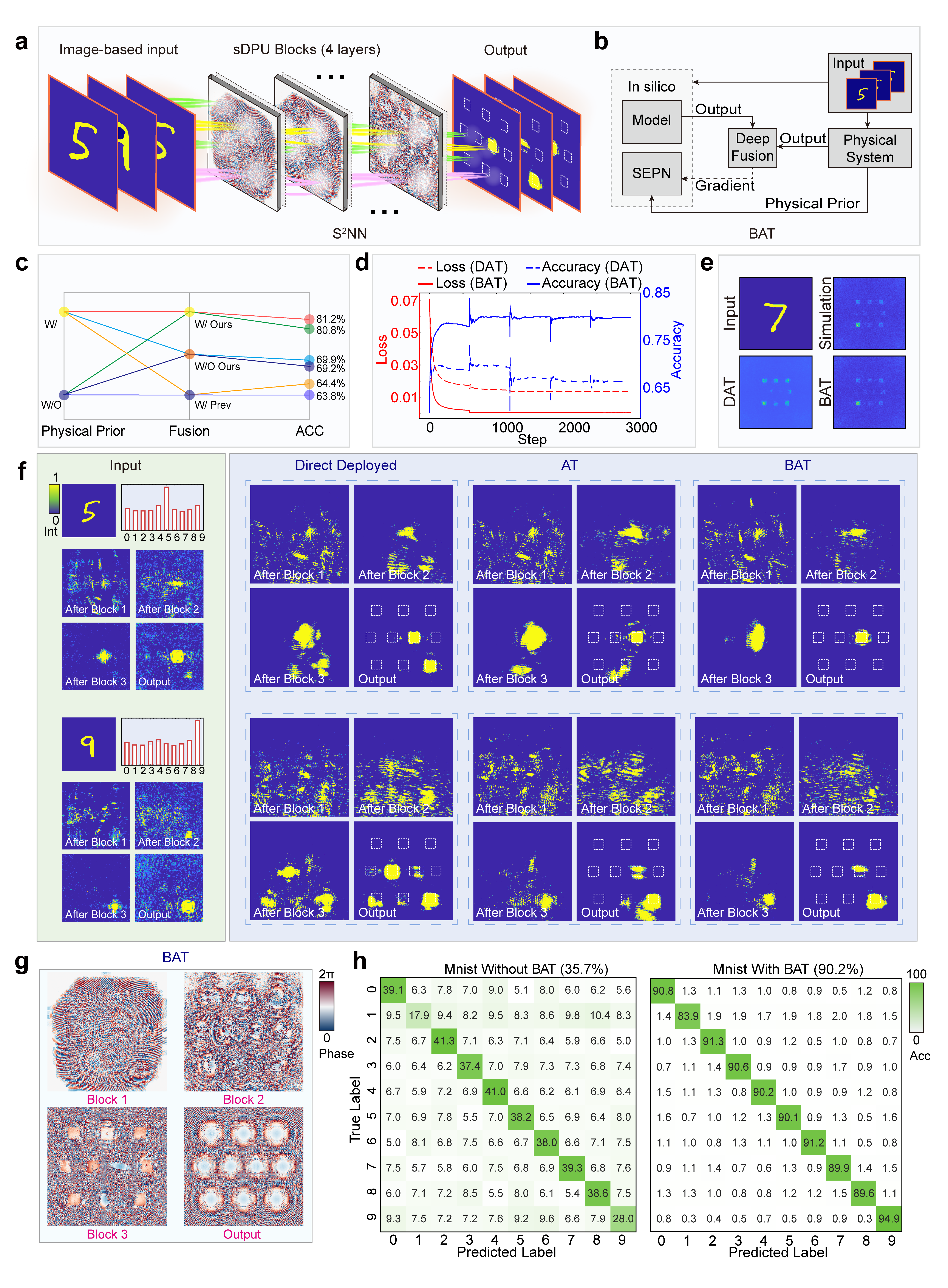}
\vspace{-0.5cm}
\caption{\textbf{Architecture and performance of S\textsuperscript{2}NN.~}\textbf{a,} S\textsuperscript{2}NN architecture: Multi-layers in S\textsuperscript{2}NN can be achieved by reusing the sDPU. 
\textbf{b,} BAT: BAT introduces physical priors and deep fusion to improve the physical system fidelity.
\textbf{c,} Ablation study of physical prior and fusion for ours BAT method compared to the DAT.
\textbf{d,} The loss plots and comparisons between BAT and DAT. The SEPNs is connected after the fisrt epoch. 
\textbf{e,} Example input (up left) in S\textsuperscript{2}NN and outputs from different deployment methods. Up right: simulation output; bottom left: output after applying the DAT method; bottom right: output after applying the BAT method.
\textbf{f,} Comparison of effects between BAT, AT, and Direct Deployment of S\textsuperscript{2}NN: Two instances, the digits "5" and "9", were selected. Among the three methods, BAT shows fewer noise points and more accurate features. \textbf{g,} Visualization of phase diagrams of each layer in a 4-layer S\textsuperscript{2}NN after BAT. \textbf{h,} Confusion matrix for classifying MNIST using direct deployment and BAT-Enhanced S\textsuperscript{2}NN: BAT significantly improves accuracy during deployment.
}\label{fig2}
\end{figure}

Fig.~\ref{fig2}a shows the constructed S\textsuperscript{2}NN based on OSN. The model was trained in silico using an MNIST training set and achieved a blind-testing accuracy of 93.6\% on 10,000 digit images in a test set. The training target was set to individually map input handwritten digits, from ‘0’ to ‘9’, into ten predefined regions on the output layer (see Methods), where a classification result was determined by finding the target region with the maximum optical signal.

Fig.~\ref{fig2}b illustrates the training and inference architecture of the physical experiments, including parallel pipeline inference and binary dual adaptive training (BAT) (details can be found in Supplementary Note 4,  Note 5; Supplementary Fig. S3, S4). In BAT, we incorporate a physical prior and deep fusion into the dual adaptive training (DAT) method from Zheng et al.~\citep{bib45}, enabling the models to adapt to significant systematic errors introduced by the neuromorphic vision sensor in the OSN. The BAT consists of three parts: the experimental system, the ideal physical model, and the systematic error prediction networks (SEPNs). This process characterizes the mapping function between the ideal physical model and the experimental system by iteratively updating the network parameters of the experimental system, the ideal physical model, and the SEPNs in an end-to-end manner for each input training sample. The physically measured and numerically extracted network outputs and internal states are fused for highly accurate gradient calculation during the ideal physical model training. After completing the BAT process, the experimental system aligns better with the numerical predictions. We then used the refined system parameters in the experimental system, allowing it to adapt to significant systematic errors from various sources.

After BAT, the test dataset was fed into the inference queue with a chosen $\Delta t$ of 490$\mu s$ per frame. Spikes generated by the OSN were processed in parallel by multiple processing threads, and binary outputs from OSN were pushed to the end of the queue to achieve complete inference. This whole process for each frame cycle consumes at least 490 $\mu s$ on average, corresponding to a computing speed of up to 2036 FPS. With this entire architecture, our S\textsuperscript{2}NN network is capable of real-time processing with a response time at the microsecond level.

Furthermore, the S\textsuperscript{2}NN was experimentally tested on the benchmark handwritten digits classification task using the MNIST dataset, where BAT was applied, as illustrated schematically with four sDPU blocks in Fig.~\ref{fig2}b and shown in the system diagram in Fig.~\ref{fig1}b. Compared to the DAT method, we introduce physical prior to the training process of SEPNs and fuse the simulation states further with physical states. As shown in Fig.~\ref{fig2}c, under phase error with std of 0.28, our simulation results demonstrate superior performance of 81.2\%, an increase of 17.4\% compared to the DAT method. Following that, we computed the loss values for the DAT and BAT respectively.  The results in Fig.~\ref{fig2}d suggest that the model trained using BAT method exhibits more stable convergence. Fig.~\ref{fig2}e demonstrates that the output fields with SEPNs generated by our network closely align with the target output distribution, manifesting only minimal discrepancies. 

For an intuitive, side-by-side comparison, we simultaneously acquire experimental outputs from different physical training methods. Fig.~\ref{fig2}f presents the physical experimental results for MNIST, comparing BAT, Adaptive Training (AT)~\citep{bib57}, and direct deployment. Fig.~\ref{fig2}d shows the phase distribution of each layer after training. The test accuracies of the baseline model in an error-free system are 93.6\% with considering quantization for MNIST classification. However, due to severe systematic errors, the test accuracy decreases dramatically to 35.7\% for MNIST when directly deploying the in silico-trained model to the constructed physical system. Aside from the geometric, fabrication and quantization errors in our physical systems, at least two kinds of constantly changing errors occur at OSN which can be somewhat removed using threshold light. For the MNIST classification, AT can only improve the test accuracy from 35.7\% to 83.6\%. By contrast, the accuracy is substantially improved to 90.2\% using BAT, demonstrating the effectiveness of BAT for physical systems to adapt to system errors. The details of BAT for modelling systematic errors is further illustrated in Supplementary Note 5.

\subsection{S\textsuperscript{3}NN for video-based real-world applications}\label{subsubsec2}

To further validate the capability to recognize high-speed dynamic motion, we experimented with the KTH and Weizmann human-action video dataset using the high-performance 6-layer S\textsuperscript{3}NN (Supplementary Fig. S5). It should be noted that conventional DONN and DRNN methods significantly limit inference speed due to inherent digital processing slowdowns, information redundancy, and frequent memory operations, thus undermining the high-speed computing advantages of photonics.  Therefore, an additional pre-neural spatiotemporal processing step is implemented to effectively reduce data volume and mitigate motion blur, while simultaneously ensuring accurate recording of subtle dynamic changes. 

\begin{figure}[H]%
\centering
\includegraphics[width=0.9\textwidth]{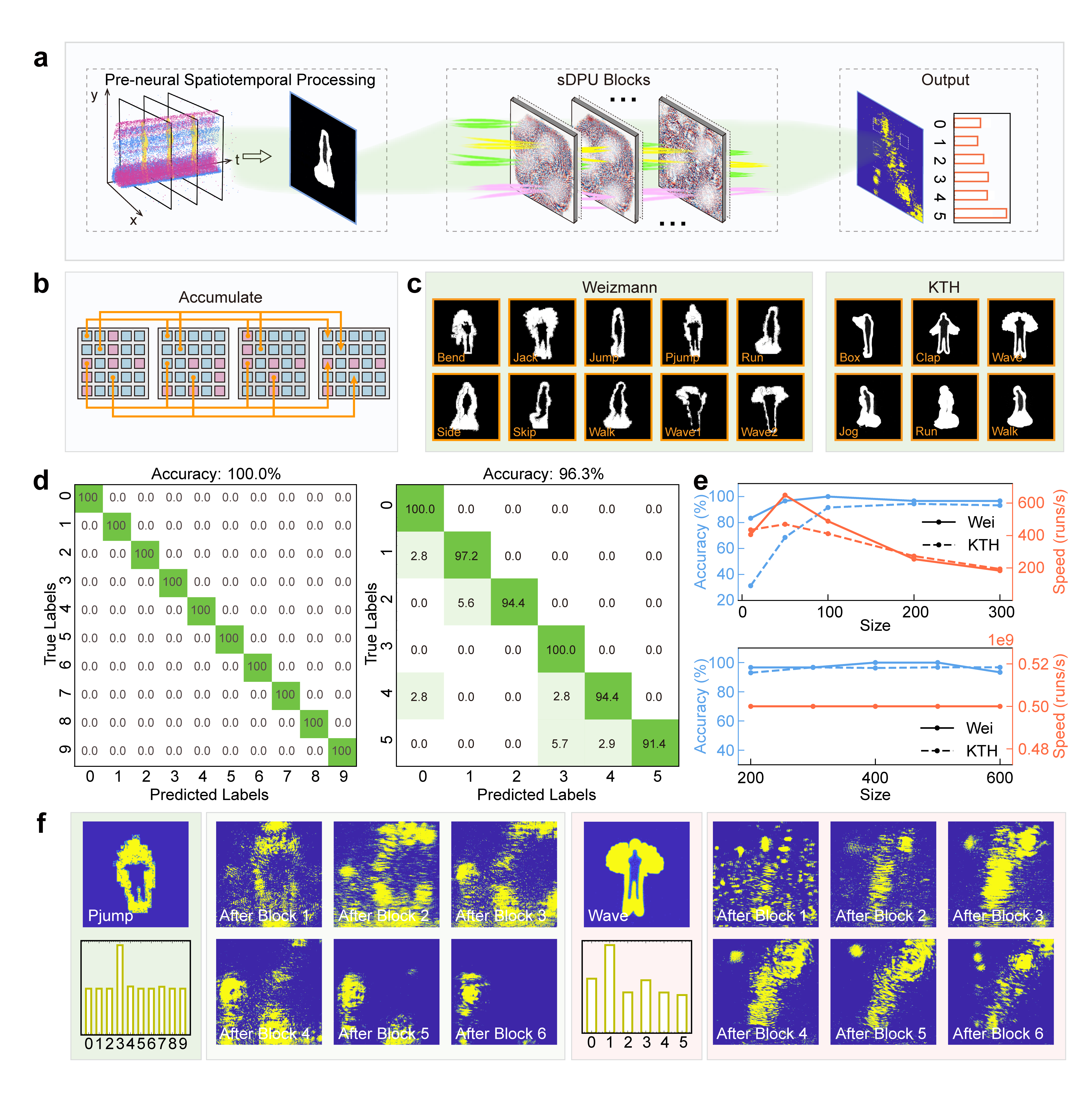}
\caption{\textbf{Architecture and performance of S\textsuperscript{3}NN.~}\textbf{a,} S\textsuperscript{3}NN architecture: Video data is converted into spatiotemporal spikes via a neuromorphic vision sensor, accumulated over time into binary frames as input to the sDPU, and after multiple layers of diffraction, the final output is produced. 
\textbf{b,} Diagram of the accumulation of spikes with a resolution of 5~$\times$~5, where red and blue denote an increase and a decrease in brightness, respectively. 
\textbf{c,} The train dataset after pre-neural spatiotemporal processing, in which each one can represent a continuous dynamic movement.  
\textbf{d,} Video accuracy confusion matrix for classifying  Weizmann (left) and KTH (right) in S\textsuperscript{3}NN. 
\textbf{e,} The impact of the final readout layer size on accuracy and recognition speed for KTH and Weizmann classification using DRNN (top) and S\textsuperscript{3}NN (bottom). In DRNN, increasing the readout layer size gradually improves accuracy, while the readout speed decreases correspondingly, leading to a reduced overall recognition speed. In S\textsuperscript{3}NN, where the readout layer is defined as the final phase layer, the size of the readout layer has almost no effect on accuracy, and since the recognition process operates at the speed of light, the recognition speed remains ultra-fast.
\textbf{f,} Experimental outputs of each layer in a 6-layer S\textsuperscript{3}NN. Left: examples from Weizmann; Right: examples from KTH.
}\label{fig3}
\end{figure}

Each pre-neural spatiotemporal computing produced a temporal spike stack that summed up the spatiotemporal dynamics over a certain time interval. The spike outputs were then compressed into binary channels. The architecture of S\textsuperscript{3}NN is shown in Fig.~\ref{fig3}a. Firstly, a neuromorphic vision sensor was used to capture dynamic actions. Based on the dimension compression approach detailed in the Methods, the spatiotemporal spike outputs from the neuromorphic vision sensor were transformed into a temporal stack within specific time intervals, which was binarized for the final inference inputs (see Supplementary Note 6 and Supplementary Fig. S6). The spikes in the temporal stack are accumulated by integrating the time dimension over a certain time interval, as illustrated in Fig.~\ref{fig3}b. The brightness of each pixel may change at various times, with red denoting an increase in brightness and blue denoting a decrease. For a certain time interval, $t_e,t_e + \Delta t$, all of the spikes are integrated into a binary frame as shown in Fig.~\ref{fig3}c. This process resulted in a series of clear, static stacks representing continuous human dynamic actions, as shown in Fig.~\ref{fig3}a-c.
 
This method also significantly reduces the amount of data required. On the one hand, the neuromorphic vision sensor reduces the amount of spike data by about 0.2\% compared with the frame-based sensor in the same scene (see Supplementary Note 6 for calculation details)~\citep{bib17}. On the other hand, when using an image to represent an spike-based video of the dynamic action, the data size will be decreased further. Thus, the pre-neural spatiotemporal processing method provides possibility for video-based real-world applications with more efficient data structure and high-speed processing ability, free from the limit of digital latency. (Details for computing performance and energy consumption for different architectures can be found in Supplementary Table S1.)

The binary temporal stacked sequences were then fed into the same physical system, which we called the temporal-stacked S\textsuperscript{2}NN (S\textsuperscript{3}NN). We compared the video accuracy of DRNN and S\textsuperscript{3}NN using the winner-takes-all policy (the action category with the most votes), where DRNN decomposed the video into 5-frame blocks for KTH and verified that DRNN could only achieve an experimental video accuracy of 91.56\%, which is lower than the 96.27\% accuracy of S\textsuperscript{3}NN for KTH. The confusion matrices of action and video classification results from the Weizmann and KTH datasets are shown in Supplementary Fig. S7 and Fig.~\ref{fig3}d, respectively. In Fig.~\ref{fig3}e, we detailed the comparison of experimental video accuracy and the read-out layer's inference speed between the DRNN and our S\textsuperscript{3}NN, noting that the electronic read-out latency would limit the overall inference speed to 300 FPS for the high-speed application. Furthermore, with at least 100k additional electronic neurons, the DRNN could maintain accuracy comparable to our S\textsuperscript{3}NN; however, it would take almost 30 times longer per inference~\citep{bib57}. Experimental testing results and corresponding inputs are shown in Fig.~\ref{fig3}f, including the correct categorized actions of “Pjump” from the Weizmann and “Wave” from the KTH dataset.

\subsection{S\textsuperscript{4}NN for single-shot classification based on optically parallel  mixture of experts}\label{subsec4}

\begin{figure}[H]%
\centering
\includegraphics[width=1.0\textwidth]{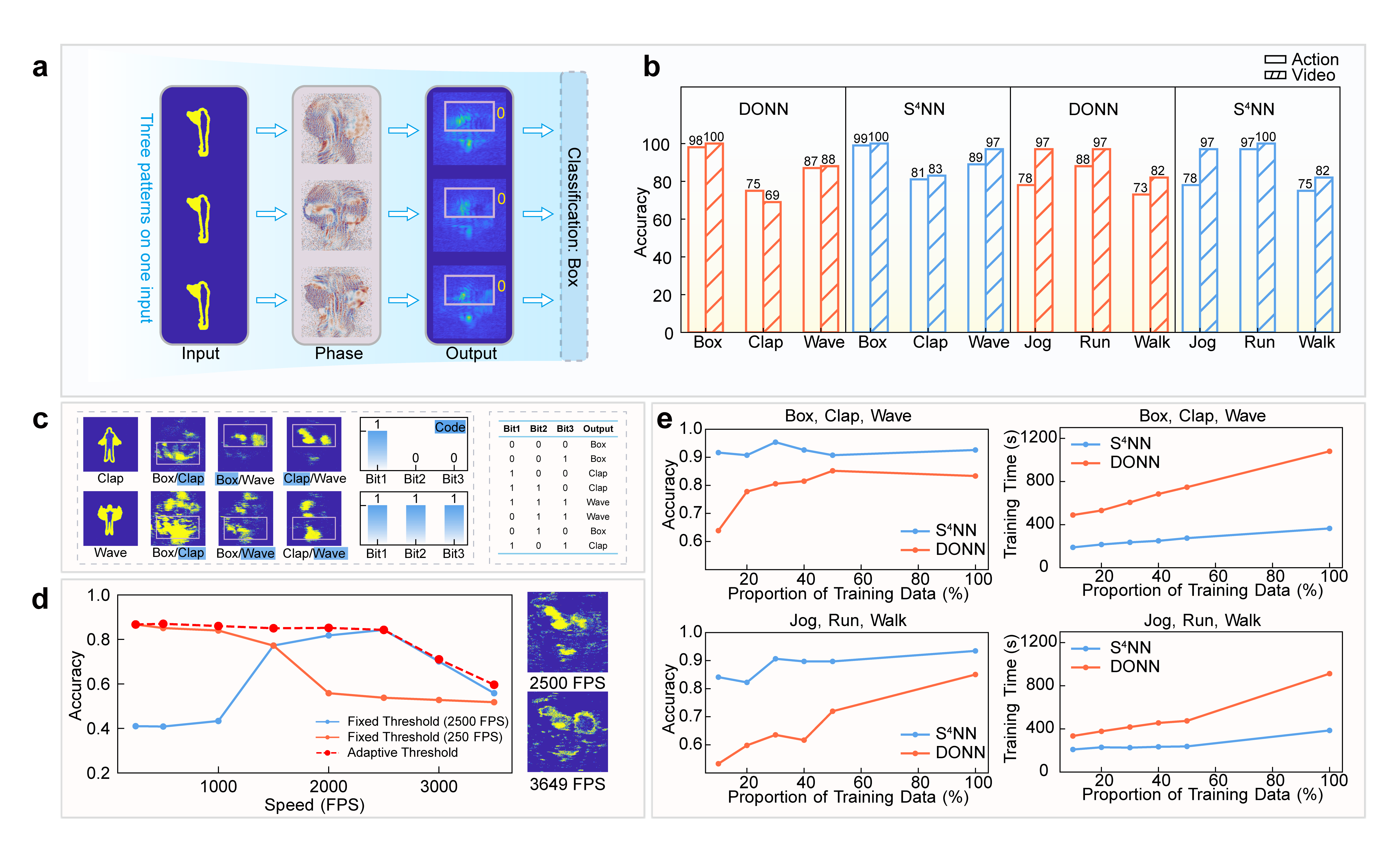}
\caption{\textbf{Architecture and performance of S\textsuperscript{4}NN.~}\textbf{a,} S\textsuperscript{4}NN architecture diagram: The optical input is duplicated into three copies, each passing through three sub-phases loaded on the SLM. The diffractive outputs are transmitted to the neuromorphic vision sensor, and the video classification results are obtained through the winner-takes-all principle. \textbf{b,} Comparison of action and video accuracy between S\textsuperscript{4}NN and ordinary DONN for each action (first and last three classes). \textbf{c,}  Experimental S\textsuperscript{4}NN outputs are shown. Left: Two examples are both from KTH.  Right: The predefined codebook provides tolerance for errors.
\textbf{d,} Left: Changes in action accuracy with variations in system operating speed, with thresholds fixed for 2500 FPS and 250 FPS.  With adaptive thresholds for operating speeds less than 2500 FPS, action accuracy is maintained at 86\% or above. Right: When operating at 2500 FPS (frames per second), the output remains complete. At 3649 FPS, the output experiences several discontinuities and missing information due to the sensor's readout latency limitation.
\textbf{e,} Comparison of changes in video accuracy and training times for S\textsuperscript{4}NN and DONN as training data volumes vary for the first three categories (Box, Clap, Wave) and last three categories (Jog, Run, Walk).
}\label{fig4}
\end{figure}

To further harness the parallelism of light, we propose using an method of OPMoE, which solves complex tasks at different levels of abstraction by leveraging the existing knowledge of multiple experts. In our case, we synthesize the outputs of individual experts specialized in subproblems via an Error-Correcting Output Codes (ECOC) coding matrix that maps each class of the multi-class problem to a binary codeword.
As shown in Fig.~\ref{fig4}a, the single-layer S\textsuperscript{4}NN includes three experts in single optical path, each responsible for distinguishing between two of the three categories. This configuration yields L binary classification results (where L is 3 in our case), which are concatenated to form a code. The generated code is then compared to the predefined codes in the codebook for each category using Hamming Distance (Hamming Distance counts the differing positions between two equal-length codes) as the metric~\citep{bib58}. The category whose code has the smallest Hamming distance from the generated code is selected as the final classification result. For instance, if the generated code is 010 (without specified classification results), it would be closest to the “Box” category code in our predefined codebook, thus demonstrating how ECOC successfully corrects potential errors in the output by leveraging the redundancy in the coding scheme. Fig.~\ref{fig4}b shows the performance of S\textsuperscript{4}NN on the task of human action recognition compared to DONN by selecting the first three (Box, Clap, Wave) and last three (Jog, Run, Walk) categories from the KTH dataset. With the OPMoE architecture, the S\textsuperscript{4}NN network successfully achieved a action/video accuracy of 90.3\%/92.6\% on the first three categories. In contrast, the single-layer DONN achieved only 83.9\%/83.3\% testing accuracy. Two specific experimental outputs and predefined codebook are shown in Fig.~\ref{fig4}c. 

Moreover, in the OPMoE method, each expert studies the binary-class problem and ultimately performs a broad range of multi-class problems, which enables the network to handle more versatile functions due to the flexibility in the number of experts and their coding strategies. Specifically, as the number of experts and code length increase, the system can tolerate errors in some binary classifiers without affecting the final result. This robustness stems from the large distances between codes of different categories and the high-accuracy of each single experts. Even if some experts in a category's code are incorrect, it could still be classified correctly if it remains closest to the original code in terms of overall distance. Our S\textsuperscript{4}NN leverages the robustness and fault tolerance of OPMoE to mitigate noise and errors inherent in optical systems, thereby improving the overall accuracy and fault tolerance of the optical neural network.

In addition, the structure of the S\textsuperscript{4}NN is appropriate for processing ultrahigh-speed visual dynamics to exploit the cut-off level of our system. 

This OPMoE approach fully utilizes the parallelism of light, as multiple binary neural networks were deployed on the SLM simultaneously, and the DMD only needed to replicate the inputs to align with the SLM. The final processing speed of the system is limited by the latency of OSN reading which consumes about 100~$\mu$s~\citep{bib47,bib49}. A more detailed analysis of the latency can be found in Supplementary Note 7.

Upon testing, the single-layer inference speed of the S\textsuperscript{4}NN reached 2500 FPS with an action accuracy of 84.3\%, and a maximum of 3649 FPS with an action accuracy of 54.5\%, as shown in Fig.~\ref{fig4}d. At this rate, the dark time ($\geq 77~\mu\text{s}$) existing in the DMD during picture switching can perform the function of a hard reset. It is important to note that the intensity of the threshold light needs to be dynamically adjusted according to the system's operating speed during deployment. As shown in Fig.~\ref{fig4}d, the threshold suitable for 250 FPS gradually becomes mismatched as the system's operating speed increases, leading to a decrease in accuracy. Similarly, the threshold suitable for 2500 FPS also experiences a mismatch when the system's operating speed is reduced. However, it can also be observed that a single threshold is applicable within an approximate range, and the required threshold light intensity is predictable as the system's operating speed changes. Therefore, this issue can be addressed later by creating a simple dynamic lookup table.

A salient advantage of OPMoE lies in their capacity to conduct efficacious training while consuming substantially fewer computational resources than previous DONN dense model counterparts. Specifically, S\textsuperscript{4}NN learns human action recognition using only 10\% of the training data, achieving less than a 1\% decrease (92\% to 91\%) in video accuracy for the action categories “Box”, “Clap”, and “Wave”, and a 9\% decrease (93\% to 84\%) for the categories “Jog”, “Run”, and “Walk”. In comparison, a single-layer and two-layer DONN exhibits a decrease of up to 19\% (83\% to 64\%) and 32\% (85\% to 53\%) in video accuracy, respectively. All in all, S\textsuperscript{4}NN requires less in silicon training time and maintains higher accuracy with reduced data usage (Fig.~\ref{fig4}e). Since classification results can be obtained with just a single layer of sDPU, its speed can be further enhanced compared to DONN.

Furthermore, by incorporating a model with OPMoE architechture, S\textsuperscript{4}NN, with only 187.5k parameters, uses significantly fewer resources than the 250k parameters of the single-layer DONN, resulting in up to a 89\% reduction in training time (Supplementary Note 8) . While the S\textsuperscript{4}NN consists of multiple small binary neural networks, its ability to exploit the parallelism of light with fewer parameters leads to superior performance (Supplementary Note 9, Supplementary Fig. S8). 

In summary, these finding reveals the potential of the S\textsuperscript{4}NN architecture in enabling rapid application via timely and efficient training, with the minimal processing time being just 274 \(\mu\)s (3649 FPS). The proposed S\textsuperscript{4}NN is 30 times faster than the previous system and is also competitive with specially optimized electronic systems.

\section{Discussion}\label{sec5}

In this study, OSN is proposed and experimentally demonstrated for high-speed dynamic machine vision. The OSN mimics biological neurons, operating in a reset-input-fire-reset cycle to provide sparse, binary outputs, fully utilizing the bandwidth and speed of optical systems. 
Our OSN-based system demonstrates remarkable versatility, seamlessly integrating into diverse model frameworks and showing immense potential for application in increasingly complex and high-speed real-world visual scenarios. We have empirically established that our OSN-based system achieves a high speed of up to 3649 FPS, which far surpasses traditional frame-based DONN systems. Notably, the OSN-based system exhibits marginally superior energy efficiency while substantially enhancing accuracy.

Our experimental results further underscore the OSN-based system's versatility and capability. Experimentally, we have successfully demonstrated high-speed applications via S\textsuperscript{2}NN, S\textsuperscript{3}NN and S\textsuperscript{4}NN in both image-based and video-based tasks. These systems achieve performance comparable to existing methods while dramatically improving speed and parameter efficiency by orders of magnitude. We demonstrate S\textsuperscript{2}NN's efficacy by its ability to classify complex dynamics of rapidly changing handwritten digits at a temporal resolution of up to 2036 FPS. Through pre-neuron processing in S\textsuperscript{3}NN, we showcase the classification of high-frequency human actions with frame intervals as brief as 490 $\mu s$. Additionally, S\textsuperscript{4}NN demonstrates high fault tolerance while operating at ultra-high speeds, processing a frame within 274 $\mu s$ at 3649 FPS.

Our OSN-based systems, leveraging its unique capability to capture brightness changes in the neuromorphic vision sensor and augmented by mimicking human neuronal activities, significantly mitigates the high information redundancy and low-speed challenges typically encountered in traditional DONN techniques. However, the high sensitivity of neuromorphic vision sensors often results in increased noise levels, presenting challenges for their applications in neural settings. To address this, our reconfigurable and unified training architecture for OSN-based systems leverages sparse characteristics and complex inherent functions to effectively reduce noise in a binary manner. By combining physical priors and deep fusion techniques in BAT, we achieve substantial improvements in both efficiency and accuracy.

The unique capabilities of the OSN enhance system operating speed and significantly reduce the data redundancy issues commonly encountered with traditional cameras. Furthermore, with advancements in high-speed spatial light modulators and the neuromorphic vision sensor, the speed of our system will further increase. Details on scalability can be found in Supplementary Note 9. Future developments in our model architecture may involve ultra-fast chemical and biological analysis, real-time autonomous driving, and high-performance large-scale optical neural information processing through reconfigurable high-speed optoelectronic computing processors.

\section{Methods}\label{sec11}
\textbf{Experimental system} 
Our physical system was constructed with commercially available optoelectronic devices to implement a single sDPU block. Complex S\textsuperscript{2}NN models such as S\textsuperscript{3}NN and S\textsuperscript{4}NN can be realized with multiple sDPU blocks by recycling. In each sDPU block, the energy source for optical computing is generated using a solid-state fibre laser (READYBeam™, FISBA AG) operating at a wavelength of 520 nm. The coherent light is polarized (LPVISC100- MP2, Thorlabs) dnf its polarization direction is changed to align with the horizontal (WPH10E-532, Thorlabs). Then, it is expanded and homogenized (GBE05-A, Thorlabs), reflected onto a DMD (V-9001c VIS, ViALUX GmbH) at the correct incident angle using a mirror (GCC-102105, Daheng Optics). The DMD consists of 2560~$\times$~1600 micro-mirrors with a pitch of 7.6~$\mu \text{m}$. These mirrors can modulate the incident light at a maximum speed of 14352 Hz. The optical field from the DMD, which is regarded as the input signal, and then optically conjugated to the phase SLM (X15223-16, HAMAMATSU) using a 4f system that comprises lenses L1 and L2 (AC508-100-A, Thorlabs). An optical iris was placed at the Fourier plane of the 4f system to filter out high-order diffraction from the DMD. After that, it passes through a 50:50 beam splitter cube (PBS251, Thorlabs), the light intensity is halved and vertically incident on the pixel plane of the phase SLM, which is a reflective liquid crystal on silicon that has a high fill factor of 96.8\% and high light utilization efficiency of 97\%. It consists of 1,272~$\times$~1,024 modulation elements, each with a size of 12.5~$\mu \text{m}$ and an 8-bit accuracy.

The outgoing light passes through the beam splitter cube again, is reflected, and then resized by lens L3 (AC508-100-A, Thorlabs). The appropriately scaled image, with the threshold light provided by a light source (HR-70-90W, Wordop), is then projected onto the pixel plane of the neuromorphic vision sensor (PEK4I36HDCD CS, PROPHESE E S.A.).

To calibrate the system, we first aligned the device pixel and obtained the effective region correspondence between the DMD and SLM, as well as that to the neuromorphic vision sensor, by projecting the chequerboard patterns and aligning them with optical translation stages. There are multiple optical components between the SLM and the neuromorphic vision sensor. Thus, we adopt the approach of adjusting the physical distance by analyzing the clarity of 1951 USAF resolution test chart hologram. The DONN system can be considered a standard diffractive imaging system. The 1951 USAF resolution test chart is used as the target image, and a corresponding phase is trained on the simulated network model according to the preset diffraction distance (30cm). This phase is then loaded onto the SLM. By adjusting the distance between the neuromorphic vision sensor and the SLM, when the maximum resolution of the 1951 USAF resolution test chart hologram is observed on the neuromorphic vision sensor, the physical distance is considered to match the simulated distance in the DONN model. In addition, the SLM was calibrated using a look-up table-calibration method to generate a linear 0–2$\pi$ phase response at a wavelength of 520 nm.

\textbf{Data processing}

\textbf{MNIST dataset}
The MNIST dataset consists of 60,000 training images and 10,000 testing images, each with a resolution of 28 × 28 pixels, categorized into one of ten classes representing the digits zero through nine. Specifically for the MNIST dataset, the initial image size of 28~$\times$~28 is first scaled up to 400~$\times$~400 and then zero-padded to 700~$\times$~700. The output of the first layer is collected and processed by the neuromorphic vision sensor, then re-input into the DMD for the second layer's inference. This process is repeated, and after the fourth layer, the classification result of the MNIST dataset is obtained.

\textbf{Weizmann dataset}
The Weizmann human action dataset comprises 90 video sequences featuring nine subjects, each performing ten distinct natural actions: bend, jumping-jack (jack), jump-forward-on-two-legs (jump), jump-in-place-on-two-legs (pjump), run, gallop-sideways (side), skip, walk, wave-one-hand (wave1), and wave-two-hands (wave2). Each sequence consists of approximately 30 to 100 frames, captured at a frame rate of 50 FPS with a resolution of 188~$\times$~144 pixels. The resolution was padded to 188~$\times$~188 pixels and further up-sampled to 400~$\times$~400 pixels, then zero-padded to 600~$\times$~600 pixels to fit the input size. Each frame is paired with a silhouette binary mask of the subject, which is used to represent different categories of human action and serves as the input to our system. For training, we utilized the actions of six subjects (60 video sequences), while the remaining three subjects (30 video sequences) were reserved for testing.

\textbf{KTH dataset}
Another used benchmark dataset for human action recognition is the KTH dataset, which consists of 600 video sequences depicting six distinct natural actions: Box, Clap, Wave, Jog, Run, and Walk. These actions are performed by 25 subjects across four different scenes. The video sequences were recorded at a frame rate of 25 FPS, resulting in approximately 350 to 600 frames per sequence, with a spatial resolution of 160~$\times$~120 pixels. 

To make the dataset compatible with our system, the silhouette binary mask of each subject in every frame was extracted using a pre-trained  DNN for semantic segmentation. Additionally, the subject was consistently aligned in each video, and frames without subjects were removed during pre-processing. Then, the resolution was padded to 160~$\times$~160 pixels and further up-sampled to 480~$\times$~480 pixels, then zero-padded to 640~$\times$~640 pixels. The model's recognition performance was assessed using a 16:9 data split.  These processed silhouette binary maskes is converted into event data by a neuromorphic vision sensor. The event data is then accumulated into event frames at fixed time intervals and input into the OSN-based system for category inference. Specifically, for an event $e$, its event stream consists of three components: timestamp ($t_e$), coordinates ($x_e$, $y_e$), and polarity ($p_e$). The polarity data is represented as 1 and 0, indicating the increase/decrease in brightness at a pixel at a given moment. By converting each positive polarity event to 1 and negative polarity event to 0 within a time period $\Delta t$, and summing up within the sensor range, an overall event distribution is obtained. By applying a threshold (set to 1) to this distribution, a pixel with a final event value greater than 1 is set to 255; otherwise, it is set to 0, resulting in a binary image.

\textbf{Neural network settings} 
Based on the forward model of the sDPU, we constructed the neural network architecture and conducted parameter optimization in silico. To balance accuracy and trainability, the phase area size ($S_{phase}$) in the single sDPU for S\textsuperscript{2}NN and S\textsuperscript{3}NN was set to 400~$\times$~400, while for S\textsuperscript{4}NN, each sub-model had a $S_{phase}$ of 250~$\times$~250, with the comparison model using the DONN architecture set to 500~$\times$~500. In addition, the DRNN for comparison was set to 400 ~$\times$~400 with a reconfigurable linear electronic fully connected read-out layer.

Consequently, the actual network parameter counts used in S\textsuperscript{2}NN and S\textsuperscript{3}NN were 640k and 960k, respectively. In the results of the comparison experiment shown in Fig.~\ref{fig4}e, the parameter counts for S\textsuperscript{4}NN's first three-class model and the last three-class model were 187.5k and 375k respectively, while the comparison model using the DONN architecture had parameter counts of 250k and 500k. In this comparison experiment, due to the different phase area usage in various models, the ratios of the input image size ($S_{input}$), propagation space size ($S_{prop}$), and $S_{phase}$ were set to be consistent during comparison ($6^2:8^2:5^2$).

During numerical modeling using the angular spectrum method, zero-padding was required at the network periphery to ensure the boundary conditions for free-space diffraction. The specific amount of zero-padding was related to the phase size.
The distance between the SLM and the sensor was optimized to 30 cm, which was derived by considering the minimum distance required to achieve full connectivity according to the Huygens-Fresnel principle and to minimize the interference between zeroth-order and first-order light during actual deployment. To facilitate fine-tuning of the phase values during BAT training and fully utilize the phase modulation range of the SLM, a 8-bit quantization was applied to constrain the phase modulation values to \(0-2\pi\). The output of each sDPU was quantized to 1-bit using the DoReFa-Net method~\citep{bib56}, with the final training objective being to maximize the energy in the label square region. The number of regions was set to match the number of labels, with the region size uniformly set to 40~$\times$~40 (S\textsuperscript{2}NN), 50~$\times$~50 (S\textsuperscript{3}NN), 125~$\times$~200 (S\textsuperscript{4}NN). Regarding the number of sDPU layers, 4 layers were used for the MNIST dataset and 6 layers for the Weizmann and KTH dataset. For the first three classes and the last three classes of the KTH dataset, 1 layer and 2 layers were used, respectively. In the actual deployment, considering that the $S_{phase}$ is 400~$\times$~400, due to the area of a single pixel on the SLM (12.5~$\times$~12.5 $\mu \text{m}^2$) being larger than that of the DMD (7.6~$\times$~7.6 $\mu \text{m}^2$) and the sensor (4.6~$\times$~4.6 $\mu \text{m}^2$), the area used on the SLM was fixed, and its actual phase side length is $12.5 \times \sqrt{S_{phase}}$ = 5000~$\mu \text{m}$. Thus, the magnification ratio of DMD should be \( \frac{5000}{7.6 \times 400} \approx 1.64 \), meaning the actual image edge length deployed on the DMD should be $1.64 \times \sqrt{S_{input}}$. 

\textbf{Training settings and BAT details in numerical experiments}
Initial models were pre-trained on a computer. The optimizer for all models was Adam, and the loss function was MSE (detailed analysis can be found in Supplementary Note 10, Supplementary Fig. S9 and S10). For the MNIST, KTH, and Weizmann datasets, the batch sizes are 20, 32, and 6 respectively, with a uniform learning rate of 0.01 and a cosine annealing scheduler adjusting the rate over a full training cycle of 50 epochs.

Subsequently, the final deployable models were obtained through the BAT method. The basic principle of BAT is to fuse the output of each physical layer with the simulation output, allowing the corrected network to learn the errors in the physical forward process. The entire process is detailed in Supplementary Note 5. Regarding specific parameters, the batch size for BAT was 10, with a learning rate of 0.001, and the number of epochs was 5, with the learning rate decaying by 0.5 every two epochs. In BAT, an additional gradient descent step was used to optimize SEPN with an initial learning rate of 0.001, decaying by 0.5 every two epochs. Furthermore, during the first few epochs of BAT, SEPN was optimized to predict systematic errors, but the connection with the physical model was severed during the optimization of the physical model, meaning SEPN did not participate in the backpropagation process or gradient calculation. After the first epoch, SEPN was roughly trained and then reconnected with the physical model for standard optimization in the last four epochs. All numerical experiments of BAT were conducted on a desktop computer with an Intel 13900K CPU at 3.90G Hz (64 cores) and an Nvidia GTX-4090 GPU with 24GB of graphics memory.

\bmhead{Supplementary information}
\indent\newline
\indent Supplementary Notes. \newline
\indent Supplementary Table S1.\newline
\indent Supplementary Figures S1 to S10.

\bmhead{Acknowledgments}
B.X. and Z.H. contributed equally to this work. This work was sponsored by Zhejiang Lab Open Research Project (NO.117008-AB2201). This work was also supported by the Guangzhou Municipal Science and Technology Project (No.2023A03J0013).
\bmhead{Competing interests}
The authors declare they have no competing interests.
\bmhead{Data and code availability}
The code will be released soon.

\bibliography{sn-article}


\begin{thebibliography}{61}
\ifx \bisbn   \undefined \def \bisbn  #1{ISBN #1}\fi
\ifx \binits  \undefined \def \binits#1{#1}\fi
\ifx \bauthor  \undefined \def \bauthor#1{#1}\fi
\ifx \batitle  \undefined \def \batitle#1{#1}\fi
\ifx \bjtitle  \undefined \def \bjtitle#1{#1}\fi
\ifx \bvolume  \undefined \def \bvolume#1{\textbf{#1}}\fi
\ifx \byear  \undefined \def \byear#1{#1}\fi
\ifx \bissue  \undefined \def \bissue#1{#1}\fi
\ifx \bfpage  \undefined \def \bfpage#1{#1}\fi
\ifx \blpage  \undefined \def \blpage #1{#1}\fi
\ifx \burl  \undefined \def \burl#1{\textsf{#1}}\fi
\ifx \doiurl  \undefined \def \doiurl#1{\url{https://doi.org/#1}}\fi
\ifx \betal  \undefined \def \betal{\textit{et al.}}\fi
\ifx \binstitute  \undefined \def \binstitute#1{#1}\fi
\ifx \binstitutionaled  \undefined \def \binstitutionaled#1{#1}\fi
\ifx \bctitle  \undefined \def \bctitle#1{#1}\fi
\ifx \beditor  \undefined \def \beditor#1{#1}\fi
\ifx \bpublisher  \undefined \def \bpublisher#1{#1}\fi
\ifx \bbtitle  \undefined \def \bbtitle#1{#1}\fi
\ifx \bedition  \undefined \def \bedition#1{#1}\fi
\ifx \bseriesno  \undefined \def \bseriesno#1{#1}\fi
\ifx \blocation  \undefined \def \blocation#1{#1}\fi
\ifx \bsertitle  \undefined \def \bsertitle#1{#1}\fi
\ifx \bsnm \undefined \def \bsnm#1{#1}\fi
\ifx \bsuffix \undefined \def \bsuffix#1{#1}\fi
\ifx \bparticle \undefined \def \bparticle#1{#1}\fi
\ifx \barticle \undefined \def \barticle#1{#1}\fi
\bibcommenthead
\ifx \bconfdate \undefined \def \bconfdate #1{#1}\fi
\ifx \botherref \undefined \def \botherref #1{#1}\fi
\ifx \url \undefined \def \url#1{\textsf{#1}}\fi
\ifx \bchapter \undefined \def \bchapter#1{#1}\fi
\ifx \bbook \undefined \def \bbook#1{#1}\fi
\ifx \bcomment \undefined \def \bcomment#1{#1}\fi
\ifx \oauthor \undefined \def \oauthor#1{#1}\fi
\ifx \citeauthoryear \undefined \def \citeauthoryear#1{#1}\fi
\ifx \endbibitem  \undefined \def \endbibitem {}\fi
\ifx \bconflocation  \undefined \def \bconflocation#1{#1}\fi
\ifx \arxivurl  \undefined \def \arxivurl#1{\textsf{#1}}\fi
\csname PreBibitemsHook\endcsname

\bibitem[\protect\citeauthoryear{Bai et~al.}{2024}]{bib1}
\begin{botherref}
\oauthor{\bsnm{Bai}, \binits{B.}},
\oauthor{\bsnm{Lee}, \binits{R.}},
\oauthor{\bsnm{Li}, \binits{Y.}},
\oauthor{\bsnm{Gan}, \binits{T.}},
\oauthor{\bsnm{Wang}, \binits{Y.}},
\oauthor{\bsnm{Jarrahi}, \binits{M.}},
\oauthor{\bsnm{Ozcan}, \binits{A.}}:
{Information-hiding cameras: Optical concealment of object information into ordinary images}.
Sci. Adv.
\textbf{10}(24)
(2024)
\doiurl{10.1126/sciadv.adn9420}
\end{botherref}
\endbibitem

\bibitem[\protect\citeauthoryear{LeCun et~al.}{2015}]{bib3}
\begin{barticle}
\bauthor{\bsnm{LeCun}, \binits{Y.}},
\bauthor{\bsnm{Bengio}, \binits{Y.}},
\bauthor{\bsnm{Hinton}, \binits{G.}}:
\batitle{{Deep learning}}.
\bjtitle{Nature}
\bvolume{521}(\bissue{7553}),
\bfpage{436}--\blpage{444}
(\byear{2015})
\doiurl{10.1038/nature14539}
\end{barticle}
\endbibitem

\bibitem[\protect\citeauthoryear{Krizhevsky et~al.}{2017}]{bib30}
\begin{barticle}
\bauthor{\bsnm{Krizhevsky}, \binits{A.}},
\bauthor{\bsnm{Sutskever}, \binits{I.}},
\bauthor{\bsnm{Hinton}, \binits{G.E.}}:
\batitle{{ImageNet classification with deep convolutional neural networks}}.
\bjtitle{Commun. ACM}
\bvolume{60}(\bissue{6}),
\bfpage{84}--\blpage{90}
(\byear{2017})
\doiurl{10.1145/3065386}
\end{barticle}
\endbibitem

\bibitem[\protect\citeauthoryear{Suo et~al.}{2023}]{bib31}
\begin{barticle}
\bauthor{\bsnm{Suo}, \binits{J.}},
\bauthor{\bsnm{Zhang}, \binits{W.}},
\bauthor{\bsnm{Gong}, \binits{J.}},
\bauthor{\bsnm{Yuan}, \binits{X.}},
\bauthor{\bsnm{Brady}, \binits{D.J.}},
\bauthor{\bsnm{Dai}, \binits{Q.}}:
\batitle{Computational imaging and artificial intelligence: The next revolution of mobile vision}.
\bjtitle{Proceedings of the IEEE}
(\byear{2023})
\doiurl{10.1109/JPROC.2023.3338272}
\end{barticle}
\endbibitem

\bibitem[\protect\citeauthoryear{Guo et~al.}{2019}]{bib48}
\begin{bchapter}
\bauthor{\bsnm{Guo}, \binits{L.}},
\bauthor{\bsnm{Lau}, \binits{J.}},
\bauthor{\bsnm{Ruan}, \binits{Z.}},
\bauthor{\bsnm{Wei}, \binits{P.}},
\bauthor{\bsnm{Cong}, \binits{J.}}:
\bctitle{Hardware acceleration of long read pairwise overlapping in genome sequencing: A race between fpga and gpu}.
In: \bbtitle{2019 IEEE 27th Annual International Symposium on Field-Programmable Custom Computing Machines (FCCM)},
pp. \bfpage{127}--\blpage{135}
(\byear{2019}).
\doiurl{10.1109/FCCM.2019.00027}
\end{bchapter}
\endbibitem

\bibitem[\protect\citeauthoryear{Mittal and Vetter}{2014}]{bib50}
\begin{barticle}
\bauthor{\bsnm{Mittal}, \binits{S.}},
\bauthor{\bsnm{Vetter}, \binits{J.S.}}:
\batitle{A survey of methods for analyzing and improving gpu energy efficiency}.
\bjtitle{ACM Computing Surveys (CSUR)}
\bvolume{47}(\bissue{2}),
\bfpage{1}--\blpage{23}
(\byear{2014})
\doiurl{10.1145/2636342}
\end{barticle}
\endbibitem

\bibitem[\protect\citeauthoryear{Br{\"u}ckerhoff-Pl{\"u}ckelmann et~al.}{2023}]{bib64}
\begin{barticle}
\bauthor{\bsnm{Br{\"u}ckerhoff-Pl{\"u}ckelmann}, \binits{F.}},
\bauthor{\bsnm{Bente}, \binits{I.}},
\bauthor{\bsnm{Becker}, \binits{M.}},
\bauthor{\bsnm{Vollmar}, \binits{N.}},
\bauthor{\bsnm{Farmakidis}, \binits{N.}},
\bauthor{\bsnm{Lomonte}, \binits{E.}},
\bauthor{\bsnm{Lenzini}, \binits{F.}},
\bauthor{\bsnm{Wright}, \binits{C.D.}},
\bauthor{\bsnm{Bhaskaran}, \binits{H.}},
\bauthor{\bsnm{Salinga}, \binits{M.}}, \betal:
\batitle{Event-driven adaptive optical neural network}.
\bjtitle{Science Advances}
\bvolume{9}(\bissue{42}),
\bfpage{9127}
(\byear{2023})
\doiurl{10.1126/sciadv.adi9127}
\end{barticle}
\endbibitem

\bibitem[\protect\citeauthoryear{Lin et~al.}{2018}]{bib2}
\begin{barticle}
\bauthor{\bsnm{Lin}, \binits{X.}},
\bauthor{\bsnm{Rivenson}, \binits{Y.}},
\bauthor{\bsnm{Yardimci}, \binits{N.T.}},
\bauthor{\bsnm{Veli}, \binits{M.}},
\bauthor{\bsnm{Luo}, \binits{Y.}},
\bauthor{\bsnm{Jarrahi}, \binits{M.}},
\bauthor{\bsnm{Ozcan}, \binits{A.}}:
\batitle{{All-optical machine learning using diffractive deep neural networks}}.
\bjtitle{Science}
\bvolume{361}(\bissue{6406}),
\bfpage{1004}--\blpage{1008}
(\byear{2018})
\doiurl{10.1126/science.aat8084}
\end{barticle}
\endbibitem

\bibitem[\protect\citeauthoryear{Jabbari and Friedman}{2023}]{bib51}
\begin{barticle}
\bauthor{\bsnm{Jabbari}, \binits{T.}},
\bauthor{\bsnm{Friedman}, \binits{E.G.}}:
\batitle{Inductive and capacitive coupling noise in superconductive vlsi circuits}.
\bjtitle{IEEE Transactions on Applied Superconductivity}
(\byear{2023})
\doiurl{10.1109/tasc.2023.3320885}
\end{barticle}
\endbibitem

\bibitem[\protect\citeauthoryear{van Nijen et~al.}{2023}]{bib52}
\begin{barticle}
\bauthor{\bsnm{Nijen}, \binits{D.A.}},
\bauthor{\bsnm{Muttillo}, \binits{M.}},
\bauthor{\bsnm{Van~Dyck}, \binits{R.}},
\bauthor{\bsnm{Poortmans}, \binits{J.}},
\bauthor{\bsnm{Zeman}, \binits{M.}},
\bauthor{\bsnm{Isabella}, \binits{O.}},
\bauthor{\bsnm{Manganiello}, \binits{P.}}:
\batitle{Revealing capacitive and inductive effects in modern industrial c-si photovoltaic cells through impedance spectroscopy}.
\bjtitle{Solar Energy Materials and Solar Cells}
\bvolume{260},
\bfpage{112486}
(\byear{2023})
\doiurl{10.1016/j.solmat.2023.112486}
\end{barticle}
\endbibitem

\bibitem[\protect\citeauthoryear{Yan et~al.}{2019}]{bib6}
\begin{barticle}
\bauthor{\bsnm{Yan}, \binits{T.}},
\bauthor{\bsnm{Wu}, \binits{J.}},
\bauthor{\bsnm{Zhou}, \binits{T.}},
\bauthor{\bsnm{Xie}, \binits{H.}},
\bauthor{\bsnm{Xu}, \binits{F.}},
\bauthor{\bsnm{Fan}, \binits{J.}},
\bauthor{\bsnm{Fang}, \binits{L.}},
\bauthor{\bsnm{Lin}, \binits{X.}},
\bauthor{\bsnm{Dai}, \binits{Q.}}:
\batitle{{Fourier-space Diffractive Deep Neural Network}}.
\bjtitle{Phys. Rev. Lett.}
\bvolume{123}(\bissue{2}),
\bfpage{023901}
(\byear{2019})
\doiurl{10.1103/PhysRevLett.123.023901}
\end{barticle}
\endbibitem

\bibitem[\protect\citeauthoryear{Zhou et~al.}{2023}]{bib7}
\begin{botherref}
\oauthor{\bsnm{Zhou}, \binits{T.}},
\oauthor{\bsnm{Wu}, \binits{W.}},
\oauthor{\bsnm{Zhang}, \binits{J.}},
\oauthor{\bsnm{Yu}, \binits{S.}},
\oauthor{\bsnm{Fang}, \binits{L.}}:
{Ultrafast dynamic machine vision with spatiotemporal photonic computing}.
Sci. Adv.
\textbf{9}(23)
(2023)
\doiurl{10.1126/sciadv.adg4391}
\end{botherref}
\endbibitem

\bibitem[\protect\citeauthoryear{Li et~al.}{2019}]{bib32}
\begin{barticle}
\bauthor{\bsnm{Li}, \binits{J.}},
\bauthor{\bsnm{Mengu}, \binits{D.}},
\bauthor{\bsnm{Luo}, \binits{Y.}},
\bauthor{\bsnm{Rivenson}, \binits{Y.}},
\bauthor{\bsnm{Ozcan}, \binits{A.}}:
\batitle{Class-specific differential detection in diffractive optical neural networks improves inference accuracy}.
\bjtitle{Advanced Photonics}
\bvolume{1}(\bissue{4}),
\bfpage{046001}
(\byear{2019})
\doiurl{10.1117/1.AP.1.4.046001}
\end{barticle}
\endbibitem

\bibitem[\protect\citeauthoryear{Chen et~al.}{2023}]{bib33}
\begin{barticle}
\bauthor{\bsnm{Chen}, \binits{Y.}},
\bauthor{\bsnm{Nazhamaiti}, \binits{M.}},
\bauthor{\bsnm{Xu}, \binits{H.}},
\bauthor{\bsnm{Meng}, \binits{Y.}},
\bauthor{\bsnm{Zhou}, \binits{T.}},
\bauthor{\bsnm{Li}, \binits{G.}},
\bauthor{\bsnm{Fan}, \binits{J.}},
\bauthor{\bsnm{Wei}, \binits{Q.}},
\bauthor{\bsnm{Wu}, \binits{J.}},
\bauthor{\bsnm{Qiao}, \binits{F.}}, \betal:
\batitle{All-analog photoelectronic chip for high-speed vision tasks}.
\bjtitle{Nature}
\bvolume{623}(\bissue{7985}),
\bfpage{48}--\blpage{57}
(\byear{2023})
\doiurl{10.1038/s41586-023-06558-8}
\end{barticle}
\endbibitem

\bibitem[\protect\citeauthoryear{Kulce et~al.}{2021}]{bib34}
\begin{barticle}
\bauthor{\bsnm{Kulce}, \binits{O.}},
\bauthor{\bsnm{Mengu}, \binits{D.}},
\bauthor{\bsnm{Rivenson}, \binits{Y.}},
\bauthor{\bsnm{Ozcan}, \binits{A.}}:
\batitle{All-optical information-processing capacity of diffractive surfaces}.
\bjtitle{Light: Science \& Applications}
\bvolume{10}(\bissue{1}),
\bfpage{25}
(\byear{2021})
\doiurl{10.1038/s41377-020-00439-9}
\end{barticle}
\endbibitem

\bibitem[\protect\citeauthoryear{Cheng et~al.}{2022}]{bib60}
\begin{barticle}
\bauthor{\bsnm{Cheng}, \binits{J.}},
\bauthor{\bsnm{Zhao}, \binits{Y.}},
\bauthor{\bsnm{Zhang}, \binits{W.}},
\bauthor{\bsnm{Zhou}, \binits{H.}},
\bauthor{\bsnm{Huang}, \binits{D.}},
\bauthor{\bsnm{Zhu}, \binits{Q.}},
\bauthor{\bsnm{Guo}, \binits{Y.}},
\bauthor{\bsnm{Xu}, \binits{B.}},
\bauthor{\bsnm{Dong}, \binits{J.}},
\bauthor{\bsnm{Zhang}, \binits{X.}}:
\batitle{A small microring array that performs large complex-valued matrix-vector multiplication}.
\bjtitle{Frontiers of Optoelectronics}
\bvolume{15}(\bissue{1}),
\bfpage{15}
(\byear{2022})
\doiurl{10.1007/s12200-022-00009-4}
\end{barticle}
\endbibitem

\bibitem[\protect\citeauthoryear{Xu et~al.}{2022}]{bib61}
\begin{barticle}
\bauthor{\bsnm{Xu}, \binits{X.}},
\bauthor{\bsnm{Ren}, \binits{G.}},
\bauthor{\bsnm{Feleppa}, \binits{T.}},
\bauthor{\bsnm{Liu}, \binits{X.}},
\bauthor{\bsnm{Boes}, \binits{A.}},
\bauthor{\bsnm{Mitchell}, \binits{A.}},
\bauthor{\bsnm{Lowery}, \binits{A.J.}}:
\batitle{Self-calibrating programmable photonic integrated circuits}.
\bjtitle{Nature Photonics}
\bvolume{16}(\bissue{8}),
\bfpage{595}--\blpage{602}
(\byear{2022})
\doiurl{10.1038/s41566-022-01020-z}
\end{barticle}
\endbibitem

\bibitem[\protect\citeauthoryear{Ashtiani et~al.}{2022}]{bib63}
\begin{barticle}
\bauthor{\bsnm{Ashtiani}, \binits{F.}},
\bauthor{\bsnm{Geers}, \binits{A.J.}},
\bauthor{\bsnm{Aflatouni}, \binits{F.}}:
\batitle{An on-chip photonic deep neural network for image classification}.
\bjtitle{Nature}
\bvolume{606}(\bissue{7914}),
\bfpage{501}--\blpage{506}
(\byear{2022})
\doiurl{10.1038/s41586-022-04714-0}
\end{barticle}
\endbibitem

\bibitem[\protect\citeauthoryear{I{\ifmmode\mbox{\c{s}}\else\c{s}\fi}{\ifmmode\imath\else\i\fi}l et~al.}{2022}]{bib8}
\begin{botherref}
\oauthor{\bsnm{I{\ifmmode\mbox{\c{s}}\else\c{s}\fi}{\ifmmode\imath\else\i\fi}l}, \binits{{\ifmmode\mbox{\c{C}}\else\c{C}\fi}.}},
\oauthor{\bsnm{Mengu}, \binits{D.}},
\oauthor{\bsnm{Zhao}, \binits{Y.}},
\oauthor{\bsnm{Tabassum}, \binits{A.}},
\oauthor{\bsnm{Li}, \binits{J.}},
\oauthor{\bsnm{Luo}, \binits{Y.}},
\oauthor{\bsnm{Jarrahi}, \binits{M.}},
\oauthor{\bsnm{Ozcan}, \binits{A.}}:
{Super-resolution image display using diffractive decoders}.
Sci. Adv.
\textbf{8}(48)
(2022)
\doiurl{10.1126/sciadv.add3433}
\end{botherref}
\endbibitem

\bibitem[\protect\citeauthoryear{Huo et~al.}{2023}]{bib9}
\begin{barticle}
\bauthor{\bsnm{Huo}, \binits{Y.}},
\bauthor{\bsnm{Bao}, \binits{H.}},
\bauthor{\bsnm{Peng}, \binits{Y.}},
\bauthor{\bsnm{Gao}, \binits{C.}},
\bauthor{\bsnm{Hua}, \binits{W.}},
\bauthor{\bsnm{Yang}, \binits{Q.}},
\bauthor{\bsnm{Li}, \binits{H.}},
\bauthor{\bsnm{Wang}, \binits{R.}},
\bauthor{\bsnm{Yoon}, \binits{S.-E.}}:
\batitle{{Optical neural network via loose neuron array and functional learning}}.
\bjtitle{Nat. Commun.}
\bvolume{14}(\bissue{2535}),
\bfpage{1}--\blpage{12}
(\byear{2023})
\doiurl{10.1038/s41467-023-37390-3}
\end{barticle}
\endbibitem

\bibitem[\protect\citeauthoryear{Cheng et~al.}{2024}]{bib10}
\begin{barticle}
\bauthor{\bsnm{Cheng}, \binits{Y.}},
\bauthor{\bsnm{Zhang}, \binits{J.}},
\bauthor{\bsnm{Zhou}, \binits{T.}},
\bauthor{\bsnm{Wang}, \binits{Y.}},
\bauthor{\bsnm{Xu}, \binits{Z.}},
\bauthor{\bsnm{Yuan}, \binits{X.}},
\bauthor{\bsnm{Fang}, \binits{L.}}:
\batitle{{Photonic neuromorphic architecture for tens-of-task lifelong learning}}.
\bjtitle{Light Sci. Appl.}
\bvolume{13}(\bissue{56}),
\bfpage{1}--\blpage{12}
(\byear{2024})
\doiurl{10.1038/s41377-024-01395-4}
\end{barticle}
\endbibitem

\bibitem[\protect\citeauthoryear{Duan et~al.}{2023}]{bib35}
\begin{barticle}
\bauthor{\bsnm{Duan}, \binits{Z.}},
\bauthor{\bsnm{Chen}, \binits{H.}},
\bauthor{\bsnm{Lin}, \binits{X.}}:
\batitle{Optical multi-task learning using multi-wavelength diffractive deep neural networks}.
\bjtitle{Nanophotonics}
\bvolume{12}(\bissue{5}),
\bfpage{893}--\blpage{903}
(\byear{2023})
\doiurl{10.1515/nanoph-2022-0615}
\end{barticle}
\endbibitem

\bibitem[\protect\citeauthoryear{Qu et~al.}{2023}]{bib36}
\begin{barticle}
\bauthor{\bsnm{Qu}, \binits{Y.}},
\bauthor{\bsnm{Lian}, \binits{H.}},
\bauthor{\bsnm{Ding}, \binits{C.}},
\bauthor{\bsnm{Liu}, \binits{H.}},
\bauthor{\bsnm{Liu}, \binits{L.}},
\bauthor{\bsnm{Yang}, \binits{J.}}:
\batitle{High-frame-rate reconfigurable diffractive neural network based on superpixels}.
\bjtitle{Optics Letters}
\bvolume{48}(\bissue{19}),
\bfpage{5025}--\blpage{5028}
(\byear{2023})
\doiurl{10.1364/OL.498712}
\end{barticle}
\endbibitem

\bibitem[\protect\citeauthoryear{Li et~al.}{2023}]{bib37}
\begin{barticle}
\bauthor{\bsnm{Li}, \binits{J.}},
\bauthor{\bsnm{Li}, \binits{X.}},
\bauthor{\bsnm{Yardimci}, \binits{N.T.}},
\bauthor{\bsnm{Hu}, \binits{J.}},
\bauthor{\bsnm{Li}, \binits{Y.}},
\bauthor{\bsnm{Chen}, \binits{J.}},
\bauthor{\bsnm{Hung}, \binits{Y.-C.}},
\bauthor{\bsnm{Jarrahi}, \binits{M.}},
\bauthor{\bsnm{Ozcan}, \binits{A.}}:
\batitle{Rapid sensing of hidden objects and defects using a single-pixel diffractive terahertz sensor}.
\bjtitle{Nature communications}
\bvolume{14}(\bissue{1}),
\bfpage{6791}
(\byear{2023})
\doiurl{10.1038/s41467-023-42554-2}
\end{barticle}
\endbibitem

\bibitem[\protect\citeauthoryear{Cheng et~al.}{2024}]{bib59}
\begin{barticle}
\bauthor{\bsnm{Cheng}, \binits{J.}},
\bauthor{\bsnm{Huang}, \binits{C.}},
\bauthor{\bsnm{Zhang}, \binits{J.}},
\bauthor{\bsnm{Wu}, \binits{B.}},
\bauthor{\bsnm{Zhang}, \binits{W.}},
\bauthor{\bsnm{Liu}, \binits{X.}},
\bauthor{\bsnm{Zhang}, \binits{J.}},
\bauthor{\bsnm{Tang}, \binits{Y.}},
\bauthor{\bsnm{Zhou}, \binits{H.}},
\bauthor{\bsnm{Zhang}, \binits{Q.}}, \betal:
\batitle{Multimodal deep learning using on-chip diffractive optics with in situ training capability}.
\bjtitle{Nature Communications}
\bvolume{15}(\bissue{1}),
\bfpage{6189}
(\byear{2024})
\doiurl{10.1038/s41467-024-50677-3}
\end{barticle}
\endbibitem

\bibitem[\protect\citeauthoryear{Milanizadeh et~al.}{2019}]{bib54}
\begin{barticle}
\bauthor{\bsnm{Milanizadeh}, \binits{M.}},
\bauthor{\bsnm{Aguiar}, \binits{D.}},
\bauthor{\bsnm{Melloni}, \binits{A.}},
\bauthor{\bsnm{Morichetti}, \binits{F.}}:
\batitle{{Canceling Thermal Cross-Talk Effects in Photonic Integrated Circuits}}.
\bjtitle{J. Lightwave Technol.}
\bvolume{37}(\bissue{4}),
\bfpage{1325}--\blpage{1332}
(\byear{2019})
\doiurl{10.1109/JLT.2019.2892512}
\end{barticle}
\endbibitem

\bibitem[\protect\citeauthoryear{Chen et~al.}{2023}]{bib53}
\begin{barticle}
\bauthor{\bsnm{Chen}, \binits{R.}},
\bauthor{\bsnm{Fang}, \binits{Z.}},
\bauthor{\bsnm{Perez}, \binits{C.}},
\bauthor{\bsnm{Miller}, \binits{F.}},
\bauthor{\bsnm{Kumari}, \binits{K.}},
\bauthor{\bsnm{Saxena}, \binits{A.}},
\bauthor{\bsnm{Zheng}, \binits{J.}},
\bauthor{\bsnm{Geiger}, \binits{S.J.}},
\bauthor{\bsnm{Goodson}, \binits{K.E.}},
\bauthor{\bsnm{Majumdar}, \binits{A.}}:
\batitle{{Non-volatile electrically programmable integrated photonics with a 5-bit operation}}.
\bjtitle{Nat. Commun.}
\bvolume{14}(\bissue{3465}),
\bfpage{1}--\blpage{10}
(\byear{2023})
\doiurl{10.1038/s41467-023-39180-3}
\end{barticle}
\endbibitem

\bibitem[\protect\citeauthoryear{Moralis-Pegios et~al.}{2024}]{bib11}
\begin{barticle}
\bauthor{\bsnm{Moralis-Pegios}, \binits{M.}},
\bauthor{\bsnm{Giamougiannis}, \binits{G.}},
\bauthor{\bsnm{Tsakyridis}, \binits{A.}},
\bauthor{\bsnm{Lazovsky}, \binits{D.}},
\bauthor{\bsnm{Pleros}, \binits{N.}}:
\batitle{{Perfect linear optics using silicon photonics}}.
\bjtitle{Nat. Commun.}
\bvolume{15}(\bissue{5468}),
\bfpage{1}--\blpage{8}
(\byear{2024})
\doiurl{10.1038/s41467-024-49768-y}
\end{barticle}
\endbibitem

\bibitem[\protect\citeauthoryear{Giamougiannis et~al.}{2023}]{bib12}
\begin{barticle}
\bauthor{\bsnm{Giamougiannis}, \binits{G.}},
\bauthor{\bsnm{Tsakyridis}, \binits{A.}},
\bauthor{\bsnm{Ma}, \binits{Y.}},
\bauthor{\bsnm{Totovi{\ifmmode\acute{c}\else\'{c}\fi}}, \binits{A.}},
\bauthor{\bsnm{Moralis-Pegios}, \binits{M.}},
\bauthor{\bsnm{Lazovsky}, \binits{D.}},
\bauthor{\bsnm{Pleros}, \binits{N.}}:
\batitle{{A Coherent Photonic Crossbar for Scalable Universal Linear Optics}}.
\bjtitle{J. Lightwave Technol.}
\bvolume{41}(\bissue{8}),
\bfpage{2425}--\blpage{2442}
(\byear{2023})
\doiurl{10.1109/JLT.2023.3234689}
\end{barticle}
\endbibitem

\bibitem[\protect\citeauthoryear{Miller}{2013}]{bib13}
\begin{barticle}
\bauthor{\bsnm{Miller}, \binits{D.A.B.}}:
\batitle{{How complicated must an optical component be?}}
\bjtitle{J. Opt. Soc. Am. A, JOSAA}
\bvolume{30}(\bissue{2}),
\bfpage{238}--\blpage{251}
(\byear{2013})
\doiurl{10.1364/JOSAA.30.000238}
\end{barticle}
\endbibitem

\bibitem[\protect\citeauthoryear{Hamerly et~al.}{2022}]{bib14}
\begin{barticle}
\bauthor{\bsnm{Hamerly}, \binits{R.}},
\bauthor{\bsnm{Bandyopadhyay}, \binits{S.}},
\bauthor{\bsnm{Englund}, \binits{D.}}:
\batitle{{Asymptotically fault-tolerant programmable photonics}}.
\bjtitle{Nat. Commun.}
\bvolume{13}(\bissue{6831}),
\bfpage{1}--\blpage{10}
(\byear{2022})
\doiurl{10.1038/s41467-022-34308-3}
\end{barticle}
\endbibitem

\bibitem[\protect\citeauthoryear{Farmakidis et~al.}{2024}]{bib38}
\begin{barticle}
\bauthor{\bsnm{Farmakidis}, \binits{N.}},
\bauthor{\bsnm{Dong}, \binits{B.}},
\bauthor{\bsnm{Bhaskaran}, \binits{H.}}:
\batitle{{Integrated photonic neuromorphic computing: opportunities and challenges}}.
\bjtitle{Nat. Rev. Electr. Eng.}
\bvolume{1}(\bissue{6}),
\bfpage{358}--\blpage{373}
(\byear{2024})
\doiurl{10.1038/s44287-024-00050-9}
\end{barticle}
\endbibitem

\bibitem[\protect\citeauthoryear{Xu et~al.}{2022}]{bib15}
\begin{barticle}
\bauthor{\bsnm{Xu}, \binits{Z.}},
\bauthor{\bsnm{Yuan}, \binits{X.}},
\bauthor{\bsnm{Zhou}, \binits{T.}},
\bauthor{\bsnm{Fang}, \binits{L.}}:
\batitle{{A multichannel optical computing architecture for advanced machine vision}}.
\bjtitle{Light Sci. Appl.}
\bvolume{11}(\bissue{255}),
\bfpage{1}--\blpage{13}
(\byear{2022})
\doiurl{10.1038/s41377-022-00945-y}
\end{barticle}
\endbibitem

\bibitem[\protect\citeauthoryear{Mengu et~al.}{2019}]{bib4}
\begin{barticle}
\bauthor{\bsnm{Mengu}, \binits{D.}},
\bauthor{\bsnm{Luo}, \binits{Y.}},
\bauthor{\bsnm{Rivenson}, \binits{Y.}},
\bauthor{\bsnm{Ozcan}, \binits{A.}}:
\batitle{{Analysis of Diffractive Optical Neural Networks and Their Integration With Electronic Neural Networks}}.
\bjtitle{IEEE J. Sel. Top. Quantum Electron.}
\bvolume{26}(\bissue{1}),
\bfpage{3700114}
(\byear{2019})
\doiurl{10.1109/JSTQE.2019.2921376}
\end{barticle}
\endbibitem

\bibitem[\protect\citeauthoryear{Wang et~al.}{2023}]{bib62}
\begin{barticle}
\bauthor{\bsnm{Wang}, \binits{T.}},
\bauthor{\bsnm{Sohoni}, \binits{M.M.}},
\bauthor{\bsnm{Wright}, \binits{L.G.}},
\bauthor{\bsnm{Stein}, \binits{M.M.}},
\bauthor{\bsnm{Ma}, \binits{S.-Y.}},
\bauthor{\bsnm{Onodera}, \binits{T.}},
\bauthor{\bsnm{Anderson}, \binits{M.G.}},
\bauthor{\bsnm{McMahon}, \binits{P.L.}}:
\batitle{Image sensing with multilayer nonlinear optical neural networks}.
\bjtitle{Nature Photonics}
\bvolume{17}(\bissue{5}),
\bfpage{408}--\blpage{415}
(\byear{2023})
\doiurl{10.1038/s41566-023-01170-8}
\end{barticle}
\endbibitem

\bibitem[\protect\citeauthoryear{Zhou et~al.}{2021}]{bib16}
\begin{barticle}
\bauthor{\bsnm{Zhou}, \binits{T.}},
\bauthor{\bsnm{Lin}, \binits{X.}},
\bauthor{\bsnm{Wu}, \binits{J.}},
\bauthor{\bsnm{Chen}, \binits{Y.}},
\bauthor{\bsnm{Xie}, \binits{H.}},
\bauthor{\bsnm{Li}, \binits{Y.}},
\bauthor{\bsnm{Fan}, \binits{J.}},
\bauthor{\bsnm{Wu}, \binits{H.}},
\bauthor{\bsnm{Fang}, \binits{L.}},
\bauthor{\bsnm{Dai}, \binits{Q.}}:
\batitle{{Large-scale neuromorphic optoelectronic computing with a reconfigurable diffractive processing unit}}.
\bjtitle{Nat. Photonics}
\bvolume{15}(\bissue{5}),
\bfpage{367}--\blpage{373}
(\byear{2021})
\doiurl{10.1038/s41566-021-00796-w}
\end{barticle}
\endbibitem

\bibitem[\protect\citeauthoryear{Li et~al.}{2024}]{bib17}
\begin{barticle}
\bauthor{\bsnm{Li}, \binits{Z.}},
\bauthor{\bsnm{Su}, \binits{H.}},
\bauthor{\bsnm{Li}, \binits{B.}},
\bauthor{\bsnm{Luan}, \binits{H.}},
\bauthor{\bsnm{Gu}, \binits{M.}},
\bauthor{\bsnm{Fang}, \binits{X.}}:
\batitle{{Event-based diffractive neural network chip for dynamic action recognition}}.
\bjtitle{Opt. Laser Technol.}
\bvolume{169},
\bfpage{110136}
(\byear{2024})
\doiurl{10.1016/j.optlastec.2023.110136}
\end{barticle}
\endbibitem

\bibitem[\protect\citeauthoryear{Mandracchia et~al.}{2024}]{bib18}
\begin{barticle}
\bauthor{\bsnm{Mandracchia}, \binits{B.}},
\bauthor{\bsnm{Zheng}, \binits{C.}},
\bauthor{\bsnm{Rajendran}, \binits{S.}},
\bauthor{\bsnm{Liu}, \binits{W.}},
\bauthor{\bsnm{Forghani}, \binits{P.}},
\bauthor{\bsnm{Xu}, \binits{C.}},
\bauthor{\bsnm{Jia}, \binits{S.}}:
\batitle{{High-speed optical imaging with sCMOS pixel reassignment}}.
\bjtitle{Nat. Commun.}
\bvolume{15}(\bissue{4598}),
\bfpage{1}--\blpage{12}
(\byear{2024})
\doiurl{10.1038/s41467-024-48987-7}
\end{barticle}
\endbibitem

\bibitem[\protect\citeauthoryear{Balderas~Silva et~al.}{2018}]{bib24}
\begin{barticle}
\bauthor{\bsnm{Balderas~Silva}, \binits{D.}},
\bauthor{\bsnm{Ponce~Cruz}, \binits{P.}},
\bauthor{\bsnm{Molina~Gutierrez}, \binits{A.}}:
\batitle{{Are the long{\textendash}short term memory and convolution neural networks really based on biological systems?}}
\bjtitle{ICT Express}
\bvolume{4}(\bissue{2}),
\bfpage{100}--\blpage{106}
(\byear{2018})
\doiurl{10.1016/j.icte.2018.04.001}
\end{barticle}
\endbibitem

\bibitem[\protect\citeauthoryear{Xia et~al.}{2024}]{bib25}
\begin{bchapter}
\bauthor{\bsnm{Xia}, \binits{W.}},
\bauthor{\bsnm{Charette}, \binits{R.}},
\bauthor{\bsnm{Oztireli}, \binits{C.}},
\bauthor{\bsnm{Xue}, \binits{J.-H.}}:
\bctitle{Dream: Visual decoding from reversing human visual system}.
In: \bbtitle{Proceedings of the IEEE/CVF Winter Conference on Applications of Computer Vision (WACV)},
pp. \bfpage{8226}--\blpage{8235}
(\byear{2024}).
\doiurl{10.1109/WACV57701.2024.00804}
\end{bchapter}
\endbibitem

\bibitem[\protect\citeauthoryear{Liu et~al.}{2020}]{bib26}
\begin{barticle}
\bauthor{\bsnm{Liu}, \binits{J.}},
\bauthor{\bsnm{Zhang}, \binits{H.}},
\bauthor{\bsnm{Yu}, \binits{T.}},
\bauthor{\bsnm{Ni}, \binits{D.}},
\bauthor{\bsnm{Ren}, \binits{L.}},
\bauthor{\bsnm{Yang}, \binits{Q.}},
\bauthor{\bsnm{Lu}, \binits{B.}},
\bauthor{\bsnm{Wang}, \binits{D.}},
\bauthor{\bsnm{Heinen}, \binits{R.}},
\bauthor{\bsnm{Axmacher}, \binits{N.}},
\bauthor{\bsnm{Xue}, \binits{G.}}:
\batitle{{Stable maintenance of multiple representational formats in human visual short-term memory}}.
\bjtitle{Proc. Natl. Acad. Sci. U.S.A.}
\bvolume{117}(\bissue{51}),
\bfpage{32329}--\blpage{32339}
(\byear{2020})
\doiurl{10.1073/pnas.2006752117}
\end{barticle}
\endbibitem

\bibitem[\protect\citeauthoryear{Guo et~al.}{2024}]{bib19}
\begin{barticle}
\bauthor{\bsnm{Guo}, \binits{R.}},
\bauthor{\bsnm{Yang}, \binits{Q.}},
\bauthor{\bsnm{Chang}, \binits{A.S.}},
\bauthor{\bsnm{Hu}, \binits{G.}},
\bauthor{\bsnm{Greene}, \binits{J.}},
\bauthor{\bsnm{Gabel}, \binits{C.V.}},
\bauthor{\bsnm{You}, \binits{S.}},
\bauthor{\bsnm{Tian}, \binits{L.}}:
\batitle{{EventLFM: event camera integrated Fourier light field microscopy for ultrafast 3D imaging}}.
\bjtitle{Light Sci. Appl.}
\bvolume{13}(\bissue{144}),
\bfpage{1}--\blpage{15}
(\byear{2024})
\doiurl{10.1038/s41377-024-01502-5}
\end{barticle}
\endbibitem

\bibitem[\protect\citeauthoryear{Wang et~al.}{2024}]{bib39}
\begin{botherref}
\oauthor{\bsnm{Wang}, \binits{J.}},
\oauthor{\bsnm{He}, \binits{J.}},
\oauthor{\bsnm{Zhang}, \binits{Z.}},
\oauthor{\bsnm{Xu}, \binits{R.}}:
Physical priors augmented event-based 3d reconstruction.
arXiv preprint
(2024)
{\href{https://arxiv.org/abs/2401.17121}{{arXiv:2401.17121}}}
\end{botherref}
\endbibitem

\bibitem[\protect\citeauthoryear{Cao et~al.}{2024}]{bib40}
\begin{bchapter}
\bauthor{\bsnm{Cao}, \binits{J.}},
\bauthor{\bsnm{Zheng}, \binits{X.}},
\bauthor{\bsnm{Lyu}, \binits{Y.}},
\bauthor{\bsnm{Wang}, \binits{J.}},
\bauthor{\bsnm{Xu}, \binits{R.}},
\bauthor{\bsnm{Wang}, \binits{L.}}:
\bctitle{Chasing day and night: Towards robust and efficient all-day object detection guided by an event camera}.
In: \bbtitle{2024 IEEE International Conference on Robotics and Automation (ICRA)},
pp. \bfpage{9026}--\blpage{9032}
(\byear{2024}).
\doiurl{10.48550/arXiv.2309.09297} .
\bcomment{IEEE}
\end{bchapter}
\endbibitem

\bibitem[\protect\citeauthoryear{Cox et~al.}{2020}]{bib49}
\begin{bchapter}
\bauthor{\bsnm{Cox}, \binits{J.}},
\bauthor{\bsnm{Ashok}, \binits{A.}},
\bauthor{\bsnm{Morley}, \binits{N.}}:
\bctitle{An analysis framework for event-based sensor performance}.
In: \bbtitle{Unconventional Imaging and Adaptive Optics 2020},
vol. \bseriesno{11508},
pp. \bfpage{63}--\blpage{78}
(\byear{2020}).
\doiurl{10.1117/12.2567620} .
\bcomment{SPIE}
\end{bchapter}
\endbibitem

\bibitem[\protect\citeauthoryear{Cabriel et~al.}{2023}]{bib20}
\begin{barticle}
\bauthor{\bsnm{Cabriel}, \binits{C.}},
\bauthor{\bsnm{Monfort}, \binits{T.}},
\bauthor{\bsnm{Specht}, \binits{C.G.}},
\bauthor{\bsnm{Izeddin}, \binits{I.}}:
\batitle{{Event-based vision sensor for fast and dense single-molecule localization microscopy}}.
\bjtitle{Nat. Photonics}
\bvolume{17}(\bissue{12}),
\bfpage{1105}--\blpage{1113}
(\byear{2023})
\doiurl{10.1038/s41566-023-01308-8}
\end{barticle}
\endbibitem

\bibitem[\protect\citeauthoryear{Yang et~al.}{2024}]{bib21}
\begin{barticle}
\bauthor{\bsnm{Yang}, \binits{Z.}},
\bauthor{\bsnm{Wang}, \binits{T.}},
\bauthor{\bsnm{Lin}, \binits{Y.}},
\bauthor{\bsnm{Chen}, \binits{Y.}},
\bauthor{\bsnm{Zeng}, \binits{H.}},
\bauthor{\bsnm{Pei}, \binits{J.}},
\bauthor{\bsnm{Wang}, \binits{J.}},
\bauthor{\bsnm{Liu}, \binits{X.}},
\bauthor{\bsnm{Zhou}, \binits{Y.}},
\bauthor{\bsnm{Zhang}, \binits{J.}},
\bauthor{\bsnm{Wang}, \binits{X.}},
\bauthor{\bsnm{Lv}, \binits{X.}},
\bauthor{\bsnm{Zhao}, \binits{R.}},
\bauthor{\bsnm{Shi}, \binits{L.}}:
\batitle{{A vision chip with complementary pathways for open-world sensing}}.
\bjtitle{Nature}
\bvolume{629}(\bissue{8014}),
\bfpage{1027}--\blpage{1033}
(\byear{2024})
\doiurl{10.1038/s41586-024-07358-4}
\end{barticle}
\endbibitem

\bibitem[\protect\citeauthoryear{Freeman et~al.}{2024}]{bib22}
\begin{bchapter}
\bauthor{\bsnm{Freeman}, \binits{A.C.}},
\bauthor{\bsnm{Mayer-Patel}, \binits{K.}},
\bauthor{\bsnm{Singh}, \binits{M.}}:
\bctitle{Accelerated event-based feature detection and compression for surveillance video systems}.
In: \bbtitle{Proceedings of the 15th ACM Multimedia Systems Conference}.
\bsertitle{MMSys '24},
pp. \bfpage{132}--\blpage{143}.
\bpublisher{Association for Computing Machinery},
\blocation{New York, NY, USA}
(\byear{2024}).
\doiurl{10.1145/3625468.3647618}
\end{bchapter}
\endbibitem

\bibitem[\protect\citeauthoryear{Klenk et~al.}{2023}]{bib23}
\begin{barticle}
\bauthor{\bsnm{Klenk}, \binits{S.}},
\bauthor{\bsnm{Koestler}, \binits{L.}},
\bauthor{\bsnm{Scaramuzza}, \binits{D.}},
\bauthor{\bsnm{Cremers}, \binits{D.}}:
\batitle{{E-NeRF: Neural Radiance Fields From a Moving Event Camera}}.
\bjtitle{IEEE Rob. Autom. Lett.}
\bvolume{8}(\bissue{3}),
\bfpage{1587}--\blpage{1594}
(\byear{2023})
\doiurl{10.1109/LRA.2023.3240646}
\end{barticle}
\endbibitem

\bibitem[\protect\citeauthoryear{Lecun et~al.}{1998}]{bib27}
\begin{barticle}
\bauthor{\bsnm{Lecun}, \binits{Y.}},
\bauthor{\bsnm{Bottou}, \binits{L.}},
\bauthor{\bsnm{Bengio}, \binits{Y.}},
\bauthor{\bsnm{Haffner}, \binits{P.}}:
\batitle{{Gradient-based learning applied to document recognition}}.
\bjtitle{Proc. IEEE}
\bvolume{86}(\bissue{11}),
\bfpage{2278}--\blpage{2324}
(\byear{1998})
\doiurl{10.1109/5.726791}
\end{barticle}
\endbibitem

\bibitem[\protect\citeauthoryear{Schuldt et~al.}{2004}]{bib28}
\begin{bchapter}
\bauthor{\bsnm{Schuldt}, \binits{C.}},
\bauthor{\bsnm{Laptev}, \binits{I.}},
\bauthor{\bsnm{Caputo}, \binits{B.}}:
\bctitle{Recognizing human actions: a local svm approach}.
In: \bbtitle{Proceedings of the 17th International Conference on Pattern Recognition, 2004. ICPR 2004.},
vol. \bseriesno{3},
pp. \bfpage{32}--\blpage{36}
(\byear{2004}).
\doiurl{10.1109/ICPR.2004.1334462}
\end{bchapter}
\endbibitem

\bibitem[\protect\citeauthoryear{Blank et~al.}{2005}]{bib29}
\begin{bchapter}
\bauthor{\bsnm{Blank}, \binits{M.}},
\bauthor{\bsnm{Gorelick}, \binits{L.}},
\bauthor{\bsnm{Shechtman}, \binits{E.}},
\bauthor{\bsnm{Irani}, \binits{M.}},
\bauthor{\bsnm{Basri}, \binits{R.}}:
\bctitle{Actions as space-time shapes}.
In: \bbtitle{Tenth IEEE International Conference on Computer Vision (ICCV'05) Volume 1},
vol. \bseriesno{2},
pp. \bfpage{1395}--\blpage{14022}
(\byear{2005}).
\doiurl{10.1109/ICCV.2005.28}
\end{bchapter}
\endbibitem

\bibitem[\protect\citeauthoryear{Krizhevsky et~al.}{2017}]{bib41}
\begin{barticle}
\bauthor{\bsnm{Krizhevsky}, \binits{A.}},
\bauthor{\bsnm{Sutskever}, \binits{I.}},
\bauthor{\bsnm{Hinton}, \binits{G.E.}}:
\batitle{Imagenet classification with deep convolutional neural networks}.
\bjtitle{Communications of the ACM}
\bvolume{60}(\bissue{6}),
\bfpage{84}--\blpage{90}
(\byear{2017})
\doiurl{10.1145/3065386}
\end{barticle}
\endbibitem

\bibitem[\protect\citeauthoryear{Desislavov et~al.}{2023}]{bib42}
\begin{barticle}
\bauthor{\bsnm{Desislavov}, \binits{R.}},
\bauthor{\bsnm{Mart{\'\i}nez-Plumed}, \binits{F.}},
\bauthor{\bsnm{Hern{\'a}ndez-Orallo}, \binits{J.}}:
\batitle{Trends in ai inference energy consumption: Beyond the performance-vs-parameter laws of deep learning}.
\bjtitle{Sustainable Computing: Informatics and Systems}
\bvolume{38},
\bfpage{100857}
(\byear{2023})
\doiurl{10.1016/j.suscom.2023.100857}
\end{barticle}
\endbibitem

\bibitem[\protect\citeauthoryear{Liu et~al.}{2024}]{bib44}
\begin{barticle}
\bauthor{\bsnm{Liu}, \binits{G.-T.}},
\bauthor{\bsnm{Shen}, \binits{Y.-W.}},
\bauthor{\bsnm{Li}, \binits{R.-Q.}},
\bauthor{\bsnm{Yu}, \binits{J.}},
\bauthor{\bsnm{He}, \binits{X.}},
\bauthor{\bsnm{Wang}, \binits{C.}},
\bauthor{\bsnm{Wang}, \binits{C.}}:
\batitle{{Optical ReLU-like activation function based on a semiconductor laser with optical injection}}.
\bjtitle{Opt. Lett.}
\bvolume{49}(\bissue{4}),
\bfpage{818}--\blpage{821}
(\byear{2024})
\doiurl{10.1364/OL.511113}
\end{barticle}
\endbibitem

\bibitem[\protect\citeauthoryear{Gallego et~al.}{2020}]{bib43}
\begin{barticle}
\bauthor{\bsnm{Gallego}, \binits{G.}},
\bauthor{\bsnm{Delbr{\"u}ck}, \binits{T.}},
\bauthor{\bsnm{Orchard}, \binits{G.}},
\bauthor{\bsnm{Bartolozzi}, \binits{C.}},
\bauthor{\bsnm{Taba}, \binits{B.}},
\bauthor{\bsnm{Censi}, \binits{A.}},
\bauthor{\bsnm{Leutenegger}, \binits{S.}},
\bauthor{\bsnm{Davison}, \binits{A.J.}},
\bauthor{\bsnm{Conradt}, \binits{J.}},
\bauthor{\bsnm{Daniilidis}, \binits{K.}}, \betal:
\batitle{Event-based vision: A survey}.
\bjtitle{IEEE transactions on pattern analysis and machine intelligence}
\bvolume{44}(\bissue{1}),
\bfpage{154}--\blpage{180}
(\byear{2020})
\doiurl{10.1109/TPAMI.2020.3008413}
\end{barticle}
\endbibitem

\bibitem[\protect\citeauthoryear{Zheng et~al.}{2023}]{bib45}
\begin{barticle}
\bauthor{\bsnm{Zheng}, \binits{Z.}},
\bauthor{\bsnm{Duan}, \binits{Z.}},
\bauthor{\bsnm{Chen}, \binits{H.}},
\bauthor{\bsnm{Yang}, \binits{R.}},
\bauthor{\bsnm{Gao}, \binits{S.}},
\bauthor{\bsnm{Zhang}, \binits{H.}},
\bauthor{\bsnm{Xiong}, \binits{H.}},
\bauthor{\bsnm{Lin}, \binits{X.}}:
\batitle{{Dual adaptive training of photonic neural networks}}.
\bjtitle{Nat. Mach. Intell.}
\bvolume{5}(\bissue{10}),
\bfpage{1119}--\blpage{1129}
(\byear{2023})
\doiurl{10.1038/s42256-023-00723-4}
\end{barticle}
\endbibitem

\bibitem[\protect\citeauthoryear{Zhou et~al.}{2021}]{bib57}
\begin{barticle}
\bauthor{\bsnm{Zhou}, \binits{T.}},
\bauthor{\bsnm{Lin}, \binits{X.}},
\bauthor{\bsnm{Wu}, \binits{J.}},
\bauthor{\bsnm{Chen}, \binits{Y.}},
\bauthor{\bsnm{Xie}, \binits{H.}},
\bauthor{\bsnm{Li}, \binits{Y.}},
\bauthor{\bsnm{Fan}, \binits{J.}},
\bauthor{\bsnm{Wu}, \binits{H.}},
\bauthor{\bsnm{Fang}, \binits{L.}},
\bauthor{\bsnm{Dai}, \binits{Q.}}:
\batitle{Large-scale neuromorphic optoelectronic computing with a reconfigurable diffractive processing unit}.
\bjtitle{Nature Photonics}
\bvolume{15}(\bissue{5}),
\bfpage{367}--\blpage{373}
(\byear{2021})
\doiurl{10.1038/s41566-021-00796-w}
\end{barticle}
\endbibitem

\bibitem[\protect\citeauthoryear{Hamming}{1950}]{bib58}
\begin{barticle}
\bauthor{\bsnm{Hamming}, \binits{R.W.}}:
\batitle{Error detecting and error correcting codes}.
\bjtitle{The Bell System Technical Journal}
\bvolume{29}(\bissue{2}),
\bfpage{147}--\blpage{160}
(\byear{1950})
\doiurl{10.1002/j.1538-7305.1950.tb00463.x}
\end{barticle}
\endbibitem

\bibitem[\protect\citeauthoryear{Haessig et~al.}{2021}]{bib47}
\begin{botherref}
\oauthor{\bsnm{Haessig}, \binits{G.}},
\oauthor{\bsnm{Joubert}, \binits{D.}},
\oauthor{\bsnm{Haque}, \binits{J.}},
\oauthor{\bsnm{Chen}, \binits{Y.}},
\oauthor{\bsnm{Milde}, \binits{M.}},
\oauthor{\bsnm{Delbruck}, \binits{T.}},
\oauthor{\bsnm{Gruev}, \binits{V.}}:
Bio-inspired polarization event camera.
arXiv preprint
(2021)
{\href{https://arxiv.org/abs/2112.01933}{{arXiv:2112.01933}}}
\end{botherref}
\endbibitem

\bibitem[\protect\citeauthoryear{Zhou et~al.}{2016}]{bib56}
\begin{botherref}
\oauthor{\bsnm{Zhou}, \binits{S.}},
\oauthor{\bsnm{Wu}, \binits{Y.}},
\oauthor{\bsnm{Ni}, \binits{Z.}},
\oauthor{\bsnm{Zhou}, \binits{X.}},
\oauthor{\bsnm{Wen}, \binits{H.}},
\oauthor{\bsnm{Zou}, \binits{Y.}}:
Dorefa-net: Training low bitwidth convolutional neural networks with low bitwidth gradients.
arXiv preprint
(2016)
{\href{https://arxiv.org/abs/1606.06160}{{arXiv:1606.06160}}}
\end{botherref}
\endbibitem

\end{thebibliography}

\end{document}